\documentclass[12pt,a4paper,english]{article}
\usepackage[T1]{fontenc}
\usepackage[utf8]{inputenc}
\usepackage{amsmath}
\usepackage{amssymb}
\usepackage{graphicx}
\usepackage{esint}

\makeatletter

\providecommand{\tabularnewline}{\\}

\numberwithin{equation}{section}
\numberwithin{figure}{section}
\numberwithin{table}{section}

\usepackage{a4wide}

\makeatother

\usepackage{babel}
\begin{document}

\title{\begin{flushright}{\normalsize ITP-Budapest Report No. 656}\end{flushright}\vspace{1cm}Sine-Gordon
multi-soliton form factors in finite volume}

\author{G. Z. Fehér$^{1}$, T. Pálmai$^{2}$ and G. Takács$^{3}$\\
~\\
$^{1}$\emph{Eötvös University, Budapest}\\
$^{2}$\emph{Budapest University of Technology and Economics}\\
$^{3}$\emph{HAS Theoretical Physics Research Group, }\\
\emph{Eötvös University, Budapest}}

\date{25th February 2012\emph{}\\
\emph{~}\\
\emph{This work is dedicated to the memory of Zalán Horváth (1943-2011).}}
\maketitle
\begin{abstract}
Multi-soliton form factors in sine-Gordon theory from the bootstrap
are compared to finite volume matrix elements computed using the truncated
conformal space approach. We find convincing agreement, and resolve
most of the issues raised in a previous work.
\end{abstract}

\section{Introduction}

The matrix elements of local operators (form factors) are central
objects in quantum field theory. In two-dimensional integrable quantum
field theory the $S$ matrix can be obtained exactly in the framework
of factorized scattering developed in \cite{zam-zam} (for a later
review see \cite{Mussardo:1992uc}). It was shown in \cite{Karowski:1978vz}
that in such theories using the scattering amplitudes as input it
is possible to obtain a set of equations satisfied by the form factors.
The complete system of form factor equations, which provides the basis
for a programmatic approach (the so-called form factor bootstrap)
was proposed in \cite{Kirillov:1987jp}. For a detailed and thorough
exposition of the subject we refer to \cite{Smirnov:1992vz}; later
this approach was also extended to form factors of boundary operators
\cite{Bajnok:2006ze,Takacs:2008je}. 

Although the connection with the Lagrangian formulation of quantum
field theory is rather indirect in the bootstrap approach, it is thought
that the general solution of the form factor axioms determines the
complete local operator algebra of the theory. This expectation was
confirmed in many cases by explicit comparison of the space of solutions
to the spectrum of local operators as described by the ultraviolet
limiting conformal field theory \cite{Cardy:1990pc,Koubek:1993ke,Koubek:1994di,Koubek:1994gk,Koubek:1994zp,Smirnov:1995jp,Delfino:1995zk,Delfino:2007bt};
the mathematical foundation is provided by the local commutativity
theorem stating that operators specified by solutions of the form
factor bootstrap are mutually local \cite{Smirnov:1992vz}. Another
important piece of information comes from correlation functions. In
the framework of quantum field theory, the operator matrix elements
can be used to build a spectral representation for the correlation
functions which provides a large distance expansion; this idea was
implemented in integrable models using form factors obtained from
the bootstrap in \cite{Yurov:1990kv}. On the other hand, the Lagrangian
or perturbed conformal field theory formulation allows one to obtain
a short-distance expansion, which can then be compared provided there
is an overlap between their regimes of validity \cite{Zamolodchikov:1990bk}.
Other evidence for the correspondence between the field theory and
the solutions of the form factor bootstrap results from evaluating
sum rules like Zamolodchikov's $c$-theorem \cite{Zamolodchikov:1986gt}
or the $\Delta$-theorem \cite{Delfino:1996nf}, both of which can
be used to express conformal data as spectral sums in terms of form
factors. Direct comparisons with multi-particle matrix elements are
not so readily available, except for perturbative or $1/N$ calculations
in some simple cases \cite{Karowski:1978vz}. 

In this paper we study form factors in finite volume, based on the
approach developed in \cite{Pozsgay:2007kn,Pozsgay:2007gx}. Finite
volume form factors have also been studied in other approaches \cite{Smirnov:1998kv,korepin-slavnov,Mussardo:2003ji};
in addition, finite temperature form factors \cite{Doyon:2006pv}
are also related to this problem, as finite temperature is equivalent
to compactified Euclidean time, and thus to a finite volume setting. 

One of the advantages of the framework developed in \cite{Pozsgay:2007kn,Pozsgay:2007gx}
is that it allows for a direct comparison of solutions of the form
factor axioms to field theory dynamics. This program has been successfully
pursued in the case of diagonal scattering theories (those without
particle mass degeneracies), both in the bulk and with boundary \cite{Pozsgay:2007kn,Pozsgay:2007gx,Pozsgay:2009pv,Kormos:2007qx}.
However, an extension to theories with non-diagonal scattering is
still missing. The first steps were taken in \cite{Feher:2011aa}
with a study of sine-Gordon breather and two-soliton form factors;
later the effect of exponential corrections (more specifically so-called
$\mu$-terms) was also studied in detail \cite{Takacs:2011nb}. Even
earlier, finite volume breather form factors were already used in
studying resonances \cite{Pozsgay:2006wb} and form factor perturbation
theory \cite{Takacs:2009fu}. The present work is a natural continuation
of this line of research, substantially extending and improving upon
the previous results. Previously, there has been no way to study multi-soliton
form factors because their integral representations could not be numerically
evaluated. This was made possible by a quite involved and tedious
numerical construction; the details of this technique are reported
elsewhere \cite{Palmai:2011nb}. 

It is important to realize that non-diagonal theories, whose spectra
contain some nontrivial particle multiplets (typically organized into
representations of some group symmetry), such as sine-Gordon or the
$O(3)$ nonlinear sigma model are very important for condensed matter
applications (e.g. to spin chains; cf. \cite{Essler:2004ht}). The
finite volume description of form factors can be used to develop a
low-temperature and large-distance expansion for finite-temperature
correlation functions \cite{Pozsgay:2007gx,Essler:2009zz,Pozsgay:2010cr},
which could in turn be used to explain experimental data, e.g. from
inelastic neutron scattering \cite{Essler:2007jp,Essler:2009zz}.
Another interesting application of finite volume form factors is the
computation of one-point functions of bulk operators on a strip with
integrable boundary conditions: the approach developed in \cite{Kormos:2010ae}
is in principle valid for general (i.e. non-diagonal) scattering.

Therefore the extension to non-diagonal theories is an interesting
direction. Sine-Gordon model can be considered as the prototype of
a non-diagonal scattering theory, and it has the advantage that its
finite volume spectra and form factors can be studied numerically
using the truncated conformal space approach, originally developed
by Yurov and Zamolodchikov for the scaling Lee-Yang model \cite{Yurov:1989yu},
but later extended to the sine-Gordon theory \cite{Feverati:1998va}.
Its exact form factors are also known in full generality \cite{smirnov_ff,Lukyanov:1993pn,Lukyanov:1997bp,Babujian:1998uw,Babujian:2001xn},
and so it is a useful playground to test our theoretical ideas on
finite volume form factors. 

To summarize, the motivations of the present work are:
\begin{itemize}
\item To continue extending the description of finite volume form factors,
initiated in \cite{Pozsgay:2007kn,Pozsgay:2007gx}, to theories with
non-diagonal scattering. 
\item We perform the first detailed direct test of the sine-Gordon multi-soliton
form factors (conjectured from the bootstrap) along the lines of \cite{Pozsgay:2007kn,Pozsgay:2007gx},
i.e. by comparing them directly to numerically determined matrix elements
computed from solving the explicit field theory dynamics in finite
volume.
\item We also wish to make sure that the numerical representation developed
in \cite{Palmai:2011nb} are correct. These numerical results are
intended to be used later for several independent lines of research,
so testing and refining them is important.
\item In the previous work \cite{Feher:2011aa}, some issues were left unresolved.
These were related to a sign observed in diagonal one-soliton matrix
elements, and a numerical discrepancy in the comparison of diagonal
matrix elements. Here we solve the first problem and present evidence
that the second one is related to truncation errors inherent in the
TCSA method.
\end{itemize}
The paper is organized as follows. After a brief review of the necessary
facts about sine-Gordon model in section \ref{sec:Brief-review-of},
we recall the formalism for finite volume soliton form factors in
section \ref{sec:Soliton-form-factors-in-finite-volume}. Using the
formalism developed in \cite{Feher:2011aa}, we give theoretical predictions
for finite volume matrix elements between multi-soliton states in
section \ref{sec:Soliton-form-factors-in-finite-volume}, which are
compared to numerical data from the truncated conformal space approach
in section \ref{sec:Numerical-results}. Section \ref{sec:Conclusions-and-outlook}
is devoted to the conclusions and outlines remaining problems, to
be investigated further.

\section{Brief review of sine-Gordon model\label{sec:Brief-review-of}}

\subsection{Action and $S$ matrix\label{sub:Action-and-}}

The classical action of the theory is 
\[
\mathcal{A}=\int d^{2}x\left(\frac{1}{2}\partial_{\mu}\Phi\partial^{\mu}\Phi+\frac{m_{0}^{2}}{\beta^{2}}\cos\beta\Phi\right)
\]
The fundamental excitations are a doublet of soliton/antisoliton of
mass $M$. Their exact $S$ matrix can be written as \cite{zam-zam}
\begin{equation}
\mathcal{S}_{i_{1}i_{2}}^{j_{1}j_{2}}(\theta,\xi)=S_{i_{1}i_{2}}^{j_{1}j_{2}}(\theta,\xi)S_{0}(\theta,\xi)\label{eq:sg_smatrix}
\end{equation}
where
\begin{eqnarray*}
 &  & S_{++}^{++}(\theta,\xi)=S_{--}^{--}(\theta,\xi)=1\\
 &  & S_{+-}^{+-}(\theta,\xi)=S_{-+}^{-+}(\theta,\xi)=S_{T}(\theta,\xi)\\
 &  & S_{+-}^{-+}(\theta,\xi)=S_{-+}^{+-}(\theta,\xi)=S_{R}(\theta,\xi)
\end{eqnarray*}
and
\begin{eqnarray*}
 &  & S_{T}(\theta,\xi)=\frac{\sinh\left(\frac{\theta}{\xi}\right)}{\sinh\left(\frac{i\pi-\theta}{\xi}\right)}\qquad,\qquad S_{R}(\theta,\xi)=\frac{i\sin\left(\frac{\pi}{\xi}\right)}{\sinh\left(\frac{i\pi-\theta}{\xi}\right)}\\
 &  & S_{0}(\theta,\xi)=-\exp\left\{ -i\int_{0}^{\infty}\frac{dt}{t}\frac{\sinh\frac{\pi(1-\xi)t}{2}}{\sinh\frac{\pi\xi t}{2}\cosh\frac{\pi t}{2}}\sin\theta t\right\} \\
 &  & =-\left(\prod_{k=1}^{n}\frac{ik\pi\xi+\theta}{ik\pi\xi-\theta}\right)\exp\Bigg\{-i\int_{0}^{\infty}\frac{dt}{t}\sin\theta t\\
 &  & \quad\times\frac{\left[2\sinh\frac{\pi(1-\xi)t}{2}\mathrm{e}^{-n\pi\xi t}+\left(\mathrm{e}^{-n\pi\xi t}-1\right)\left(\mathrm{e}^{\pi(\xi-1)t/2}+\mathrm{e}^{-\pi(1+\xi)t/2}\right)\right]}{2\sinh\frac{\pi\xi t}{2}\cosh\frac{\pi t}{2}}\Bigg\}
\end{eqnarray*}
(the latter representation is valid for any value of $n\in\mathbb{N}$
and makes the integral representation converge faster and further
away from the real $\theta$ axis). Besides the solitons, the spectrum
of theory contains also breathers; we omit details since these play
no role in the sequel. We also introduced the parameter
\[
\xi=\frac{\beta^{2}}{8\pi-\beta^{2}}
\]
Another representation of the theory is as a free massless boson conformal
field theory (CFT) perturbed by a relevant operator. The Hamiltonian
can be written as
\begin{equation}
H=\int dx\frac{1}{2}:\left(\partial_{t}\Phi\right)^{2}+\left(\partial_{x}\Phi\right)^{2}:+\mu\int dx:\cos\beta\Phi:\label{eq:pcft_action}
\end{equation}
where the semicolon denotes normal ordering in terms of the modes
of the $\mu=0$ massless field. In this case, due to anomalous dimension
of the normal ordered cosine operator, the coupling constant $\mu$
has dimension
\[
\mu\sim\left[\mbox{mass}\right]^{2-\beta^{2}/4\pi}
\]
so it defines the mass scale of the model and the dimensionless coupling
parameter is $\beta$.

\subsection{Soliton form factors\label{sub:Soliton-form-factors}}

We consider only exponentials of the bosonic field $\Phi$. Their
vacuum expectation value is known exactly \cite{Lukyanov:1996jj}:
\begin{eqnarray}
\mathcal{G}_{a}(\beta)=\langle\mathrm{e}^{ia\beta\Phi}\rangle & = & \left[\frac{M\sqrt{\pi}\Gamma\left(\frac{4\pi}{8\pi-\beta^{2}}\right)}{2\Gamma\left(\frac{\beta^{2}/2}{8\pi-\beta^{2}}\right)}\right]^{\frac{a^{2}\beta^{2}}{4\pi}}\exp\Bigg\{\int_{0}^{\infty}\frac{dt}{t}\Bigg[-\frac{a^{2}\beta^{2}}{4\pi}e^{-2t}\nonumber \\
 &  & +\frac{\sinh^{2}\left(\frac{a}{4\pi}t\right)}{2\sinh\left(\frac{\beta^{2}}{8\pi}t\right)\cosh\left(\left(1-\frac{\beta^{2}}{8\pi}\right)t\right)\sinh t}\Bigg]\Bigg\}\label{eq:exactvev}
\end{eqnarray}
with $M$ denoting the soliton mass related to the coupling $\mu$
defined in via \cite{Zamolodchikov:1995xk} 
\begin{equation}
\mu=\frac{2\Gamma(\Delta)}{\pi\Gamma(1-\Delta)}\left(\frac{\sqrt{\pi}\Gamma\left(\frac{1}{2-2\Delta}\right)M}{2\Gamma\left(\frac{\Delta}{2-2\Delta}\right)}\right)^{2-2\Delta}\qquad,\qquad\Delta=\frac{\beta^{2}}{8\pi}\label{eq:mass_scale}
\end{equation}
As for multi-soliton form factors, at present there are three independent
constructions: the earliest one by Smirnov (reviewed in \cite{smirnov_ff}),
the free field representation by Lukyanov \cite{Lukyanov:1993pn,Lukyanov:1997bp}
and the work by Babujian et al. \cite{Babujian:1998uw,Babujian:2001xn}.
Here we use formulae from Lukyanov's work \cite{Lukyanov:1997bp};
however, certain of his conventions are different and therefore we
change the labeling of the form factors accordingly (see eqn. (\ref{eq:fromlukyanovstoours})
below). The reason is that the form factors we use satisfy form factor
bootstrap relations which are slightly different from Lukyanov's conventions;
in this we conform to the conventions of the papers \cite{Pozsgay:2007kn,Pozsgay:2007gx}.
In our notations, the form factor equations are:

I. Lorentz-invariance
\begin{equation}
F_{i_{1}\dots i_{N}}^{\mathcal{O}}(\theta_{1}+\Lambda,\dots,\theta_{N}+\Lambda)=\mathrm{e}^{s(\mathcal{O})\Lambda}F_{i_{1}\dots i_{N}}^{\mathcal{O}}(\theta_{1},\dots,\theta_{N})\label{eq:Lorentzaxiom}
\end{equation}
where $s(\mathcal{O})$ is the Lorentz spin of the operator $\mathcal{O}$.

II. Exchange:

\begin{center}
\begin{eqnarray}
 &  & F_{i_{1}\dots i_{k}i_{k+1}\dots i_{N}}^{\mathcal{O}}(\theta_{1},\dots,\theta_{k},\theta_{k+1},\dots,\theta_{N})=\nonumber \\
 &  & \qquad S_{i_{k}i_{k+1}}^{j_{k}j_{k+1}}(\theta_{k}-\theta_{k+1})F_{i_{1}\dots j_{k+1}j_{k}\dots i_{N}}^{\mathcal{O}}(\theta_{1},\dots,\theta_{k+1},\theta_{k},\dots,\theta_{N})\label{eq:exchangeaxiom}
\end{eqnarray}

\par\end{center}

III. Cyclic permutation: 
\begin{equation}
F_{i_{1}i_{2}\dots i_{N}}^{\mathcal{O}}(\theta_{1}+2i\pi,\theta_{2},\dots,\theta_{N})=\mathrm{e}^{2\pi i\omega(\mathcal{O})}F_{i_{2}\dots i_{N}i_{1}}^{\mathcal{O}}(\theta_{2},\dots,\theta_{N},\theta_{1})\label{eq:cyclicaxiom}
\end{equation}
where $\omega(\mathcal{O})$ is the mutual locality index between
the operator $\mathcal{O}$ and the asymptotic field that creates
the solitons.

IV. Kinematical singularity
\begin{eqnarray}
 &  & -i\mathop{\textrm{Res}}_{\theta=\theta^{'}}F_{i\, k\, i_{1}\dots i_{N}}^{\mathcal{O}}(\theta+i\pi,\theta^{'},\theta_{1},\dots,\theta_{n})=\label{eq:kinematicalaxiom}\\
 &  & \qquad C_{ik'}\left(\delta_{k}^{k'}-\mathrm{e}^{2\pi i\omega(\mathcal{O})}S_{ki_{1}}^{k_{1}j_{1}}(\theta'-\theta_{1})S_{k_{1}i_{2}}^{k_{2}j_{2}}(\theta'-\theta_{2})\dots S_{k_{n-1}i_{n}}^{k'j_{n}}(\theta'-\theta_{N})\right)F_{j_{1}\dots j_{N}}^{\mathcal{O}}(\theta_{1},\dots,\theta_{N})\nonumber 
\end{eqnarray}
where $C$ is the charge conjugation matrix.

V. Dynamical singularity 
\begin{equation}
-i\mathop{\textrm{Res}}_{\epsilon=0}F_{i\, j\, i_{1}\dots i_{N}}^{\mathcal{O}}(\theta+i\bar{u}_{jk}^{i}/2+\epsilon,\theta^{'}-i\bar{u}_{ik}^{j}/2,\theta_{1},\dots,\theta_{N})=\Gamma_{ij}^{k}F_{k\, i_{1}\dots i_{N}}^{\mathcal{O}}(\theta,\theta_{1},\dots,\theta_{N})\label{eq:dynamicalaxiom}
\end{equation}
whenever $k$ occurs as the bound state of the particles $i$ and
$j$, corresponding to a bound state pole of the $S$ matrix, where
$\Gamma_{ij}^{k}$ is the on-shell three-particle coupling and $u_{ij}^{k}$
is the so-called fusion angle. The fusion angles satisfy
\begin{eqnarray*}
m_{k}^{2} & = & m_{i}^{2}+m_{j}^{2}+2m_{i}m_{j}\cos u_{ij}^{k}\\
2\pi & = & u_{ij}^{k}+u_{ik}^{j}+u_{jk}^{i}
\end{eqnarray*}
and we also used the notation $\bar{u}_{ij}^{k}=\pi-u_{ij}^{k}$.
Equations I-V are supplemented by the assumption of maximum analyticity
(i.e. that the form factors are meromorphic functions which only have
the singularities prescribed by the axioms) and possible further conditions
expressing properties of the particular operator whose form factors
are sought.

The form factors of the operator 
\[
\mathcal{O}_{a}=\mathrm{e}^{ia\beta\Phi}
\]
which satisfy equations (\ref{eq:Lorentzaxiom}-\ref{eq:kinematicalaxiom})
with the locality index
\[
\omega(\mathcal{O}_{a})=a\bmod1
\]
can be obtained from 
\begin{align}
F_{\sigma_{1}\dots\sigma_{2n}}^{a}(\theta_{1},\dots,\theta_{2n}) & =(-1)^{n}\mathcal{F}_{-\sigma_{2N}\dots-\sigma_{1}}^{(a)}(\theta_{2n},\dots,\theta_{1})\nonumber \\
 & =(-1)^{n}\mathcal{F}_{\sigma_{2n}\dots\sigma_{1}}^{(-a)}(\theta_{2n},\dots,\theta_{1})\label{eq:fromlukyanovstoours}
\end{align}
where the functions $\mathcal{F}$ (derived by Lukyanov) are specified
in appendix \ref{sec:Explicit-formulae-for-FF}. Equation (\ref{eq:dynamicalaxiom})
for the dynamical singularities can then be used to construct form
factors of breathers, which are bound states of a soliton with an
antisoliton.

Note that the sign factor $(-1)^{n}$ corresponds to a redefinition
of the relative phase between a soliton and an antisoliton. In our
previous work \cite{Feher:2011aa}, it was noticed that such a sign
was necessary for a full agreement between the finite size data and
the theoretical predictions. Since then we realized that this is explained
by the difference between the conventions used for the form factor
equations between Lukyanov's work \cite{Lukyanov:1997bp} and the
finite volume form factor formalism developed in \cite{Pozsgay:2007kn,Pozsgay:2007gx}.
Similarly, the other sign change is related to another difference
in the conventions, namely the sign of the sine-Gordon field $\Phi$,
which can be compensated by either flipping the sign of $a$ or exchanging
the soliton with the antisoliton (charge conjugation). Finally, the
rapidity ordering is again a matter of convention, this time that
of fixing the basis for the asymptotic multi-particle states.

\section{Soliton form factors in finite volume \label{sec:Soliton-form-factors-in-finite-volume}}

\subsection{Finite volume form factors in non-diagonal theories\label{sub:Finite-volume-form}}

The formulae for finite volume form factors, derived in \cite{Pozsgay:2007kn,Pozsgay:2007gx},
were generalized for the case of non-diagonal theories in \cite{Feher:2011aa}.
Here we only recall the necessary facts; for more details the reader
is referred to the original papers. 

In finite volume $L$, the space of multi-soliton states can be labeled
by momentum quantum numbers $I_{1},\dots,I_{N}$. We introduce the
following notation for them: 
\begin{equation}
|\{I_{1},I_{2},\dots,I_{N}\}\rangle_{L}^{(r)}\label{eq:finvolstate}
\end{equation}
where the index $r$ enumerates the eigenvectors of the $n$-particle
transfer matrix, which can be written as 
\[
\mathcal{\mathcal{T}}\left(\lambda|\left\{ \theta_{1},\dots,\theta_{N}\right\} \right)_{i_{1}\dots i_{N}}^{j_{1}\dots j_{N}}=\mathcal{S}_{ai_{1}}^{c_{1}j_{1}}(\lambda-\theta_{1})\mathcal{S}_{c_{1}i_{2}}^{c_{2}j_{2}}(\lambda-\theta_{2})\dots\mathcal{S}_{c_{N-1}i_{N}}^{aj_{N}}(\lambda-\theta_{N})
\]
where $\theta_{1},\dots,\theta_{N}$ are particle rapidities. The
transfer matrix can be diagonalized simultaneously for all values
of $\lambda$:
\[
\mathcal{\mathcal{T}}\left(\lambda|\left\{ \theta_{1},\dots,\theta_{N}\right\} \right)_{i_{1}\dots i_{N}}^{j_{1}\dots j_{N}}\Psi_{j_{1}\dots j_{n}}^{(r)}\left(\left\{ \theta_{k}\right\} \right)=t^{(r)}\left(\lambda,\left\{ \theta_{k}\right\} \right)\Psi_{i_{1}\dots i_{n}}^{(r)}\left(\left\{ \theta_{k}\right\} \right)
\]
We can assume that the wave function amplitudes $\Psi^{(r)}$ are
normalized and form a complete basis:
\begin{align*}
\sum_{i_{1}\dots i_{N}}\Psi_{i_{1}\dots i_{N}}^{(r)}\left(\left\{ \theta_{k}\right\} \right)\Psi_{i_{1}\dots i_{N}}^{(s)}\left(\left\{ \theta_{k}\right\} \right)^{*} & =\delta_{rs}\\
\sum_{r}\Psi_{i_{1}\dots i_{N}}^{(r)}\left(\left\{ \theta_{k}\right\} \right)\Psi_{j_{1}\dots j_{N}}^{(r)}\left(\left\{ \theta_{k}\right\} \right)^{*} & =\delta_{i_{1}j_{1}}\dots\delta_{i_{N}j_{N}}
\end{align*}
these eigenfunctions describe the possible polarizations of the $N$
particle state with rapidities $\theta_{1},\dots,\theta_{N}$ inside
the $2^{N}$ dimensional internal space indexed by $i_{1}\dots i_{N}$.

The rapidities of the particles in the state (\ref{eq:finvolstate})
can be determined by solving the quantization conditions 
\begin{align}
Q_{j}(\theta_{1},\dots,\theta_{n}) & =ML\sinh\theta_{j}+\delta_{j}^{(r)}(\theta_{1},\dots,\theta_{N})=2\pi I_{j}\quad,\quad k=1,\dots,N\label{eq:betheyang_general}\\
 & \delta_{j}^{(r)}(\theta_{1},\dots,\theta_{N})=-i\log t^{(r)}\left(\theta_{j},\left\{ \theta_{k}\right\} \right)\nonumber 
\end{align}
When considering rapidities which solve these equations with given
quantum numbers $I_{1},\dots I_{N}$ and a specific polarization state
$r$, they will be written with a tilde as $\tilde{\theta}_{1},\dots,\tilde{\theta}_{N}$.

Using the above ingredients, the finite volume matrix elements can
then be written as 
\begin{eqnarray}
 &  & \left|\,^{(s)}\langle\{I_{1}',\dots,I_{M}'\}\vert\mathcal{O}(0,0)\vert\{I_{1},\dots,I_{N}\}\rangle_{L}^{(r)}\right|=\nonumber \\
 &  & \qquad\left|\frac{{\displaystyle F^{\mathcal{O}(s)}(\tilde{\theta}_{M}',\dots,\tilde{\theta}_{1}'|\tilde{\theta}_{1},\dots,\tilde{\theta}_{N})^{(r)}}}{\sqrt{\rho^{(r)}(\tilde{\theta}_{1},\dots,\tilde{\theta}_{N})\rho^{(s)}(\tilde{\theta}_{1}',\dots,\tilde{\theta}_{M}')}}\right|+O(\mathrm{e}^{-\mu'L})\label{eq:nondiag_genffrelation}
\end{eqnarray}
where $\rho^{(r)}$ and $\rho^{(s)}$ denote the density of states
of types $r$ and $s$,
\begin{eqnarray*}
 &  & F^{\mathcal{O}(s)}(\theta_{M}',\dots,\theta_{1}'|\theta_{1},\dots,\theta_{N})^{(r)}\\
 &  & =\sum_{j_{1}\dots j_{M}}\sum_{i_{1}\dots i_{N}}\Psi_{j_{1}\dots j_{M}}^{(s)}\left(\left\{ \theta_{k}'\right\} \right)^{*}F_{\bar{j}_{M}\dots\bar{j}_{1}i_{1}\dots i_{N}}^{\mathcal{O}}(\theta_{M}'+i\pi,\dots,\theta_{1}'+i\pi,\theta_{1},\dots,\theta_{N})\Psi_{i_{1}\dots i_{N}}^{(r)}\left(\left\{ \theta_{k}\right\} \right)
\end{eqnarray*}
and the bar denotes the antiparticle. The absolute value in (\ref{eq:nondiag_genffrelation})
and in all similar formulae below is necessary to account for the
different phase conventions of the multi-particle states used in the
form factor bootstrap and in the finite volume calculations. 

Relation (\ref{eq:nondiag_genffrelation}) is only valid for matrix
elements with no disconnected pieces, i.e. when the rapidities in
the two finite volume states are all different from each other. If
there are particles with exactly coinciding rapidities in the two
states, i.e. $\tilde{\theta}_{k}'=\tilde{\theta}_{l}$ for some $k$
and $l$, then there are further contributions. Note that equality
of two quantum numbers such as $I_{k}'=I_{l}$ is not sufficient for
the presence a disconnected contribution, as the corresponding rapidities
will in general be different due to the terms involving the phase
shifts $\delta_{j}^{(r)}$. Therefore such terms are only present
for the case when the two sets of quantum numbers are exactly identical,
and also in the special case when the two states each contain a particle
with exactly zero rapidity. At present, the disconnected terms are
only known for states with diagonal scattering; the form of these
contributions was obtained in \cite{Pozsgay:2007gx}.

\subsection{Soliton-antisoliton states}

This can be easily applied to soliton-antisoliton states. Two-soliton
states form a four dimensional space corresponding to $ss$, $s\bar{s}$,
$\bar{s}s$ and $\bar{s}\bar{s}$. Due to the charge conjugation invariance
of the $S$ matrix, the eigenvectors of the transfer matrix have definite
charge parity; together with charge conservation, this uniquely determines
them. The eigenvectors of the two-soliton transfer matrix in the neutral
subspace are \cite{Feher:2011aa}
\begin{align}
\Psi^{(+)} & =\frac{1}{\sqrt{2}}(0,+1,+1,0)\nonumber \\
\Psi^{(-)} & =\frac{1}{\sqrt{2}}(0,+1,-1,0)\label{eq:ssbarwavefun}
\end{align}
and are even/odd under charge conjugation, respectively.

This results in the following quantization conditions for the soliton-antisoliton
pair:
\begin{align}
Q_{1}^{(\pm)}(\theta_{1},\theta_{2}) & =ML\sinh\theta_{1}+\delta_{\pm}(\theta_{1}-\theta_{2})=2\pi I_{1}\nonumber \\
Q_{2}^{(\pm)}(\theta_{1},\theta_{2}) & =ML\sinh\theta_{2}+\mathcal{\delta}_{\pm}(\theta_{2}-\theta_{1})=2\pi I_{2}\label{eq:ssbarby}
\end{align}
where the phase-shifts $\delta_{\pm}$ are defined from the eigenvalues
of the two-particle $S$-matrix in the neutral subspace by 
\begin{align*}
\mathcal{S}_{+}(\theta) & =\mathcal{S}_{+-}^{+-}(\theta)+\mathcal{S}_{+-}^{-+}(\theta)=-\mathrm{e}^{i\delta_{+}(\theta)}\\
\mathcal{S}_{-}(\theta) & =\mathcal{S}_{+-}^{+-}(\theta)-\mathcal{S}_{+-}^{-+}(\theta)=\mathrm{e}^{i\delta_{-}(\theta)}
\end{align*}
and the $\pm$ distinguishes the two states (\ref{eq:ssbarwavefun}).
Note the $-$ sign introduced in the first line which ensures that
the phase-shifts are odd and continuous functions of the rapidity
$\theta$; as a consequence they vanish for $\theta=0$. Due to this
convention the $+$ states are quantized with half-integer, while
the $-$ states are quantized with integer quantum numbers.

The density of states can be written as the Jacobi determinant \cite{Feher:2011aa}
\[
\rho^{(\pm)}(\theta_{1},\theta_{2})=\left|\begin{array}{cc}
\frac{\partial Q_{1}^{(\pm)}}{\partial\theta_{1}} & \frac{\partial Q_{1}^{(\pm)}}{\partial\theta_{2}}\\
\frac{\partial Q_{2}^{(\pm)}}{\partial\theta_{1}} & \frac{\partial Q_{2}^{(\pm)}}{\partial\theta_{2}}
\end{array}\right|
\]
From (\ref{eq:nondiag_genffrelation}) we obtain
\begin{equation}
\left|\,\langle0\vert\mathcal{O}(0,0)\vert\{I_{1},I_{2}\}\rangle_{L}^{(\pm)}\right|=\frac{\left|F^{\pm}(\tilde{\theta}_{1}-\tilde{\theta}_{2})\right|}{\sqrt{\rho^{(\pm)}(\tilde{\theta}_{1},\tilde{\theta}_{2})}}+O(\mathrm{e}^{-\mu L})\label{eq:ssbar_finitevol}
\end{equation}
where 
\[
F^{\pm}(\theta)=\frac{1}{\sqrt{2}}\left(F_{+-}(\theta)\pm F_{-+}(\theta)\right)=-\frac{1}{\sqrt{2}}\left(\mathcal{F}_{-+}^{1}(\theta)\pm\mathcal{F}_{+-}^{1}(\theta)\right)
\]
in terms of (\ref{eq:twossff}) and $\tilde{\theta}_{1,2}$ are the
solutions of (\ref{eq:ssbarby}) at the given volume $L$ with quantum
numbers $I_{1,2}$. This relation was already tested in \cite{Feher:2011aa}.

Similarly one obtains
\begin{align}
\left|\,^{(s)}\langle\{I_{1}',I_{2}'\vert\mathcal{O}(0,0)\vert\{I_{1},I_{2}\}\rangle_{L}^{(r)}\right| & =\frac{\left|F^{(s)}(i\pi+\tilde{\theta}_{2}',i\pi+\tilde{\theta}_{1}',\tilde{\theta}_{1},\tilde{\theta}_{2})^{(r)}\right|}{\sqrt{\rho^{(s)}(\tilde{\theta}_{1}',\tilde{\theta}_{2}')\rho^{(r)}(\tilde{\theta}_{1},\tilde{\theta}_{2})}}+O(\mathrm{e}^{-\mu L})\nonumber \\
s,r & =\pm1\label{eq:4ffrel}
\end{align}
where 
\begin{eqnarray*}
F^{(s)}(\theta_{2}',\theta_{1}',\theta_{1},\theta_{2})^{(r)} & = & \frac{1}{2}\Bigg[F_{-++-}(\theta_{2}',\theta_{1}',\theta_{1},\theta_{2})+rF_{-+-+}(\theta_{2}',\theta_{1}',\theta_{1},\theta_{2})\\
 &  & sF_{+-+-}(\theta_{2}',\theta_{1}',\theta_{1},\theta_{2})+rsF_{+--+}(\theta_{2}',\theta_{1}',\theta_{1},\theta_{2})\Bigg]
\end{eqnarray*}
provided the matrix element is non-diagonal, i.e. the two states differ
either in their symmetry indices $s,r$, or in at least one of the
momentum quantum numbers.

States containing more than two solitons/antisolitons can be described
using the algebraic Bethe Ansatz \cite{Feher:2011aa}; we do not enter
into details as they are not needed in the sequel.

\subsection{States containing only solitons}

Another way to test the multi-soliton form factors is to consider
matrix elements where both states contain only solitons of like (say
positive) topological charge. Since their scattering is diagonal,
the formulae from \cite{Pozsgay:2007kn,Pozsgay:2007gx} are directly
applicable. The quantization relations for these states are
\begin{equation}
Q_{k}(\theta_{1},\dots,\theta_{N})=ML\sinh\theta_{k}+\sum_{l\neq k}\delta(\theta_{k}-\theta_{l})=2\pi I_{k}\quad,\quad k=1,\dots,N\label{eq:betheyang}
\end{equation}
where the phase-shift $\delta$ is defined by 
\[
\mathcal{S}_{++}^{++}(\theta)=-\mathrm{e}^{i\delta(\theta)}
\]
and the density of states is
\begin{equation}
\rho(\theta_{1},\dots,\theta_{N})_{L}=\det\mathcal{J}^{(N)}\qquad,\qquad\mathcal{J}_{kl}^{(N)}=\frac{\partial Q_{k}(\theta_{1},\dots,\theta_{N})}{\partial\theta_{l}}\quad,\quad k,l=1,\dots,N\label{eq:byjacobian}
\end{equation}
Because of the sign in the definition of the phase-shift, the $I_{k}$
are integer/half-integer for states containing on odd/even number
of solitons, respectively.

The finite volume matrix elements can be expressed as follows \cite{Pozsgay:2007kn}:
\begin{eqnarray}
 &  & \left|\langle\{I_{1}',\dots,I_{M}'\}\vert\mathcal{O}(0,0)\vert\{I_{1},\dots,I_{N}\}\rangle_{L}\right|=\nonumber \\
 &  & \qquad\frac{\left|F_{\underbrace{-\dots-}_{M}\underbrace{+\dots+}_{N}}^{\mathcal{O}}(\tilde{\theta}_{M}'+i\pi,\dots,\tilde{\theta}_{1}'+i\pi,\tilde{\theta}_{1},\dots,\tilde{\theta}_{N})\right|}{\sqrt{\rho(\tilde{\theta}_{1},\dots,\tilde{\theta}_{N})_{L}\rho(\tilde{\theta}_{1}',\dots,\tilde{\theta}_{M}')_{L}}}+O(\mathrm{e}^{-\mu L})\label{eq:genffrelation}
\end{eqnarray}
Since the topological charge of the operator 
\[
\mathcal{O}=\mathrm{e}^{i\beta\Phi}
\]
vanishes, the matrix elements are only nonzero when $N=M$. 

For the particular case of states containing only solitons we also
know the form of disconnected contributions; since their scattering
is diagonal, one can use the results from \cite{Pozsgay:2007gx}.
For diagonal matrix elements 
\begin{eqnarray}
\langle\{I_{1},\dots,I_{N}\}\vert\mathcal{O}(0,0)\vert\{I_{1},\dots,I_{N}\}\rangle_{L} & = & \frac{1}{\rho(\{1,\dots,N\})_{L}}\times\label{eq:diaggenrule}\\
 &  & \sum_{A\subset\{1,2,\dots N\}}\mathcal{F}(A)_{L}\rho(\{1,\dots,N\}\setminus A)_{L}+O(\mathrm{e}^{-\mu L})\nonumber 
\end{eqnarray}
where $|A|$ denotes the cardinal number (number of elements) of the
set $A$ 
\[
\rho(\{k_{1},\dots,k_{r}\})_{L}=\rho(\tilde{\theta}_{k_{1}},\dots,\tilde{\theta}_{k_{r}})_{L}
\]
is the $r$-particle Bethe-Yang Jacobi determinant (\ref{eq:byjacobian})
involving only the $r$-element subset $1\leq k_{1}<\dots<k_{r}\leq N$
of the $N$ particles, and
\begin{eqnarray*}
\mathcal{F}(\{k_{1},\dots,k_{r}\})_{L} & = & F_{r}^{s}(\tilde{\theta}_{k_{1}},\dots,\tilde{\theta}_{k_{r}})\\
F_{r}^{s}(\theta_{1},\dots,\theta_{l})_{i_{1}\dots i_{l}} & = & \lim_{\epsilon\rightarrow0}F^{\mathcal{O}}(\theta_{l}+i\pi+\epsilon,\dots,\theta_{1}+i\pi+\epsilon,\theta_{1},\dots,\theta_{l})_{\underbrace{-\dots-}_{r}\underbrace{+\dots+}_{r}}
\end{eqnarray*}
Besides diagonal matrix elements, the only other possibility for disconnected
terms to occur is when both states contain a stationary particle;
in our case it can only happen in matrix elements with the same (odd)
number of solitons on both sides. The general formula can be found
in \cite{Pozsgay:2007gx}; here we only quote the case needed in the
sequel:
\begin{eqnarray}
 &  & \left|\langle\{I',0,-I'\}|\mathcal{O}|\{I,0,-I\}\rangle_{L}\right|=\label{eq:zeromom}\\
 &  & \frac{\Bigg|\mathcal{F}_{1,1}(\tilde{\theta}'|\tilde{\theta})+ML\, F_{--++}^{\mathcal{O}}(i\pi+\tilde{\theta}',i\pi-\tilde{\theta}',-\tilde{\theta},\tilde{\theta})\Bigg|}{\sqrt{\rho(\tilde{\theta}',0,-\tilde{\theta}')_{L}\rho(\tilde{\theta},0,-\tilde{\theta})_{L}}}+O(\mathrm{e}^{-\mu L})\nonumber 
\end{eqnarray}
where 
\[
\mathcal{F}_{1,1}(\theta'|\theta)=\lim_{\epsilon\rightarrow0}F_{---+++}^{\mathcal{O}}(i\pi+\theta'+\epsilon,i\pi-\theta'+\epsilon,i\pi+\epsilon,0,-\theta,\theta)
\]

\section{Numerical results \label{sec:Numerical-results}}

\subsection{Numerical methods\label{sub:Numerical-methods}}

To evaluate the form factors numerically, we use the truncated conformal
space approach (TCSA) pioneered by Yurov and Zamolodchikov \cite{Yurov:1989yu}.
The extension to the sine-Gordon model was developed in \cite{Feverati:1998va}
and has found numerous applications since then. The Hilbert space
can be split by the eigenvalues of the topological charge $\mathcal{Q}$
(or winding number) and the spatial momentum $P$, where the eigenvalues
of the latter are of the form
\[
\frac{2\pi s}{L}
\]
$s$ is called the 'conformal spin'. The basis of the Hilbert space
is constructed in the ultraviolet limiting massless free boson CFT
with central charge $c=1$, and a (dimensionless) upper cutoff $E_{cut}$
is imposed on the scaling dimension (which is the sum of the left
and right conformal dimensions) of the states kept under the truncation.

In sectors with vanishing topological charge, we can make use of the
symmetry of the Hamiltonian under 
\[
\mathcal{C}:\qquad\Phi(x,t)\rightarrow-\Phi(x,t)
\]
which is equivalent to conjugation of the solitonic charge. The truncated
space can be split into $\mathcal{C}$-even and $\mathcal{C}$-odd
subspaces that have roughly equal dimensions \cite{Feher:2011aa},
which speeds up the diagonalization of the Hamiltonian by roughly
a factor of eight (the required machine time scales approximately
with the third power of matrix size). We used the program developed
for the work \cite{Feher:2011aa}, with cutoff values $E_{cut}$ ranging
between $16$ to $26$; the highest cutoff was chosen such that the
dimension does not exceed $11000$ states (in order for the program
to fit into available computer memory and also finish in a reasonable
amount of time); the maximum $E_{cut}$ permitted by this criterion
depends on the value of the sine-Gordon coupling $\beta$, and the
topological charge and spin of the sector under consideration. 

For the matrix element calculations, we chose the operator
\[
\mathcal{O}=:\mathrm{e}^{i\beta\Phi}:
\]
which is essentially one half of the interaction term in the Hamiltonian
in (\ref{eq:pcft_action}). The semicolons denote normal ordering
with respect to the $\lambda=0$ free massless boson modes. This operator
has conformal dimension
\[
\Delta_{\mathcal{O}}=\bar{\Delta}_{\mathcal{O}}=\frac{\beta^{2}}{8\pi}
\]
Using relation (\ref{eq:mass_scale}) we can express all energy levels
and matrix elements in units of (appropriate powers of) the soliton
mass $M$, and we also introduce the dimensionless volume variable
$l=ML$. The general procedure is the same as in \cite{Pozsgay:2007kn,Pozsgay:2007gx}:
the particle content of energy levels can be identified by matching
the numerical TCSA spectrum against the predictions of the Bethe-Yang
equations.

To generate the data used for comparison, altogether $342$ TCSA Hamiltonians
were diagonalized, and from them $1350$ operator matrices were computed.
Three values of couplings were used: $\xi=2/7,$ $50/239$ and $50/311$;
for each of them we evaluated the sectors $Q=0,1,2,3$ with spins
$s=0,1,2$ (for the sectors with $Q=0$, this was done separately
for the $\mathcal{C}$-even/odd projections) and with as many values
for $E_{cut}\geq16$ as the dimensionality constraint admitted. This
left us with a vast amount of useful data of which we only include
an illustrative sample; we performed the comparison for a much larger
set, with identical results to the ones presented below. 

As in all our previous works on finite volume form factors (see e.g.
\cite{Pozsgay:2007kn,Pozsgay:2007gx}), energy levels predicted by
the Bethe-Yang equation were used to identify the particle contents
of the finite volume energy levels computed numerically from the TCSA
method. Due to level crossings, at certain values of the volume $L$
there can be more than one TCSA candidate levels for a given Bethe-Yang
solution; for the data presented here we kept only unambiguously identified
levels. 

In all of the figures presented below, we denote the operator matrix
element by $f$: this means taking the absolute value of the matrix
elements which is normalized by choosing the TCSA vectors orthonormal.
This conforms to the conventions used in eqns. (\ref{eq:nondiag_genffrelation},\ref{eq:genffrelation},\ref{eq:diaggenrule},\ref{eq:zeromom}).
In all cases, the discrete points are the numerical TCSA data, while
the continuous lines are the corresponding theoretical expectations.

\subsection{Sources of deviations}

There are two sources of deviations between the theoretical predictions
and numerical results: 
\begin{enumerate}
\item Exponential finite size effects are neglected in the theoretical description
for the volume dependence of matrix elements, outlined in section
\ref{sec:Soliton-form-factors-in-finite-volume}. While they are partially
understood (especially the so-called $\mu$-terms \cite{Pozsgay:2008bf,Takacs:2011nb}),
there is no systematic description for them yet, so we do not consider
them here. Generally, they are expected to be larger for smaller $\xi$:
both because the breathers become lighter in terms of the mass scale
$M$ provided by the soliton mass (which affects so-called $F$-terms
arising from breather loops non-trivially wound around the finite
volume), and also because they become less tightly bound when considered
as bound states of other breathers (which enhances the $\mu$-terms
related to compositeness). Some of the $\mu$-terms can also be dangerously
enhanced by the analytic behaviour of form factors \cite{Takacs:2011nb},
but no sign of such behaviour was seen for the matrix element considered
in this work.
\item Truncation errors introduced by TCSA, on the other hand, generally
increase with the volume and are also larger for higher excited states.
In sine-Gordon theory, they have been observed to become smaller when
decreasing $\xi$, so the two sources of deviations behave the opposite
way when the sine-Gordon coupling is varied. Behaviour of truncation
errors in the asymptotic regime of large values of the cutoff can
be theoretically described by a Wilsonian renormalization group \cite{Feverati:2006ni,Konik:2007cb,Giokas:2011ix}.\\
Level crossings also present a problem in numerical stability, since
in their vicinity the state of interest is nearly degenerate to another
one. Since the truncation effect can be considered as an additional
perturbing operator, the level crossings are eventually lifted. However,
such a near-degeneracy greatly magnifies truncation effects on the
eigenvectors and therefore the matrix elements \cite{Kormos:2007qx}.
This is the reason behind the fact that there are some individual
numerical points that are clearly scattered away from their expected
place (cf. fig \ref{fig:q0}).
\end{enumerate}
For any quantity (energy levels and matrix elements) which is compared
between the theoretical predictions and the numerics, one can define
the ``scaling regime'', which is the volume range in which the two
sources of deviations are the smallest, i.e. the range in which truncation
errors and exponential finite size effects are approximately the same
magnitude. This range depends on the following factors:
\begin{enumerate}
\item The value of the sine-Gordon coupling $\beta$: when $\beta$ (or
equivalently) $\xi$ decreases, it shifts to larger values of the
volumes, and also becomes longer.
\item The TCSA truncation: it becomes longer when increasing the value and
also shifts to slightly higher values of the volume.
\item The quantity under considerations: as shown below, diagonal matrix
elements are the ones most affected by truncation errors, for which
we have no theoretical explanation at present.
\end{enumerate}

\subsection{\label{sub:Results-for-four-particle}Results for four-particle form
factors}

\begin{figure}
\begin{centering}
\includegraphics[width=0.85\paperwidth]{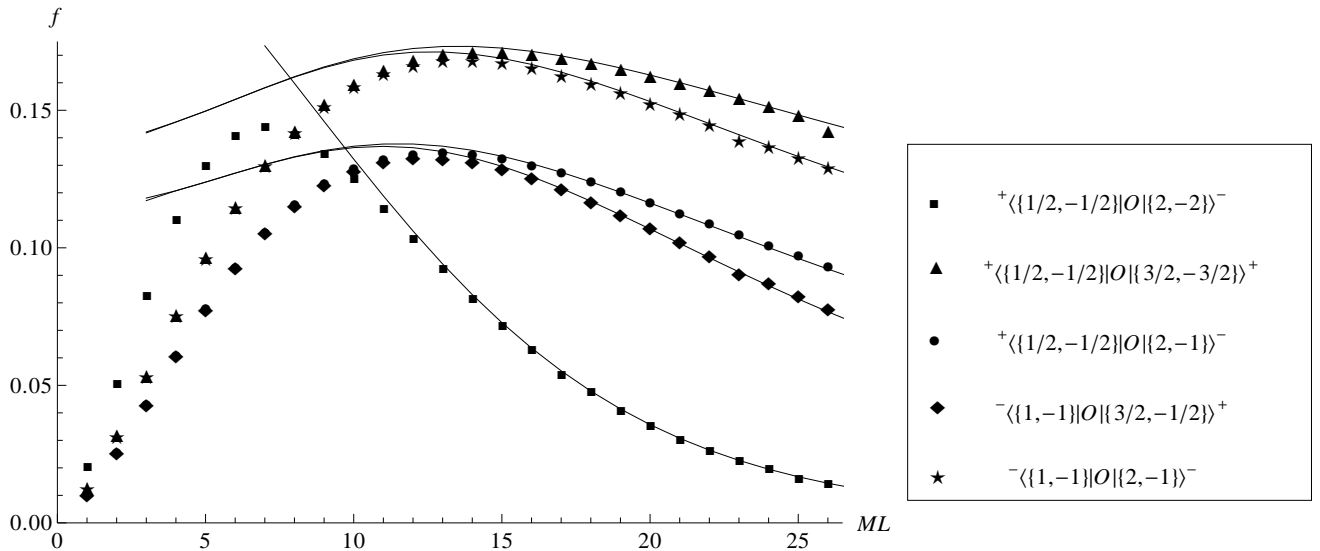}
\par\end{centering}

\caption{\label{fig:q0} Non-diagonal form factors in the $Q=0$ sector for
$\xi=50/239$. A few examples of individual data points affected by
truncation effects magnified by the vicinity of a level crossing can
be seen at $ML=23$, cf. the second and fourth lines from below (plotted
with diamonds and stars, respectively). }
\end{figure}

We can consider off-diagonal matrix elements between soliton-antisoliton
two-particle states (for the diagonal ones we do not have the theoretical
description yet, cf. the discussion in the conclusions). The theoretical
prediction is given by eqn. (\ref{eq:4ffrel}) and the comparison
is shown in figure \ref{fig:q0}. 

\begin{figure}
\begin{centering}
\includegraphics[width=0.85\paperwidth]{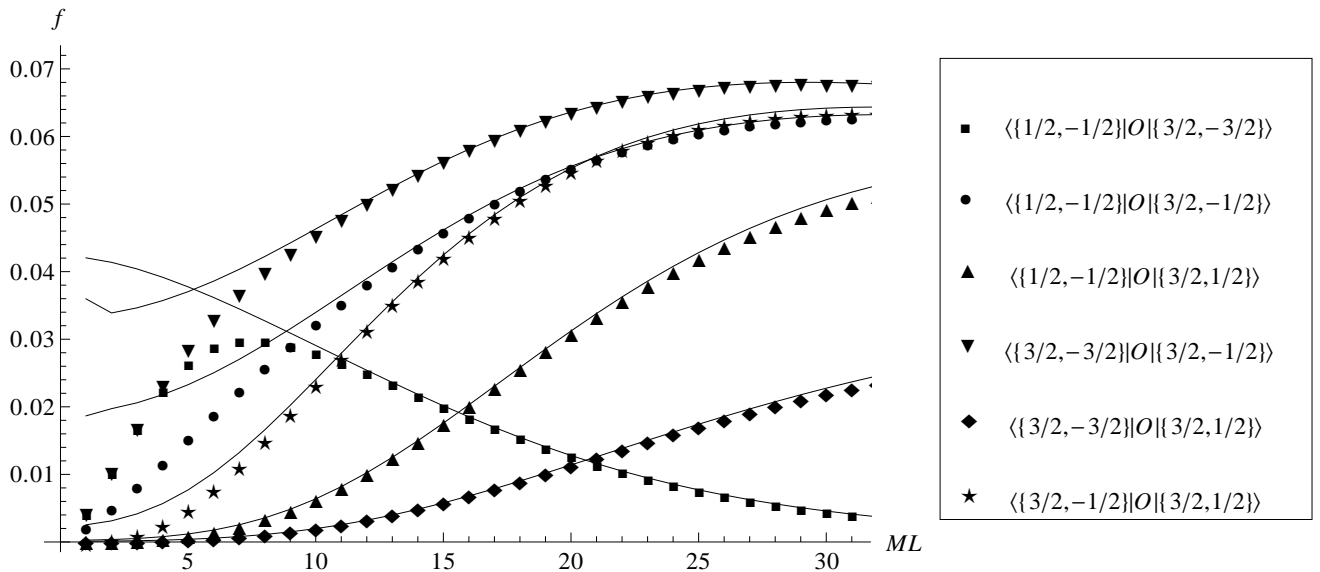}
\par\end{centering}

\caption{\label{fig:q2nondiag} Non-diagonal form factors in the $Q=2$ sector
for $\xi=50/239$}
\end{figure}

\begin{figure}
\begin{centering}
\begin{tabular}{c}
\includegraphics[width=0.65\textwidth]{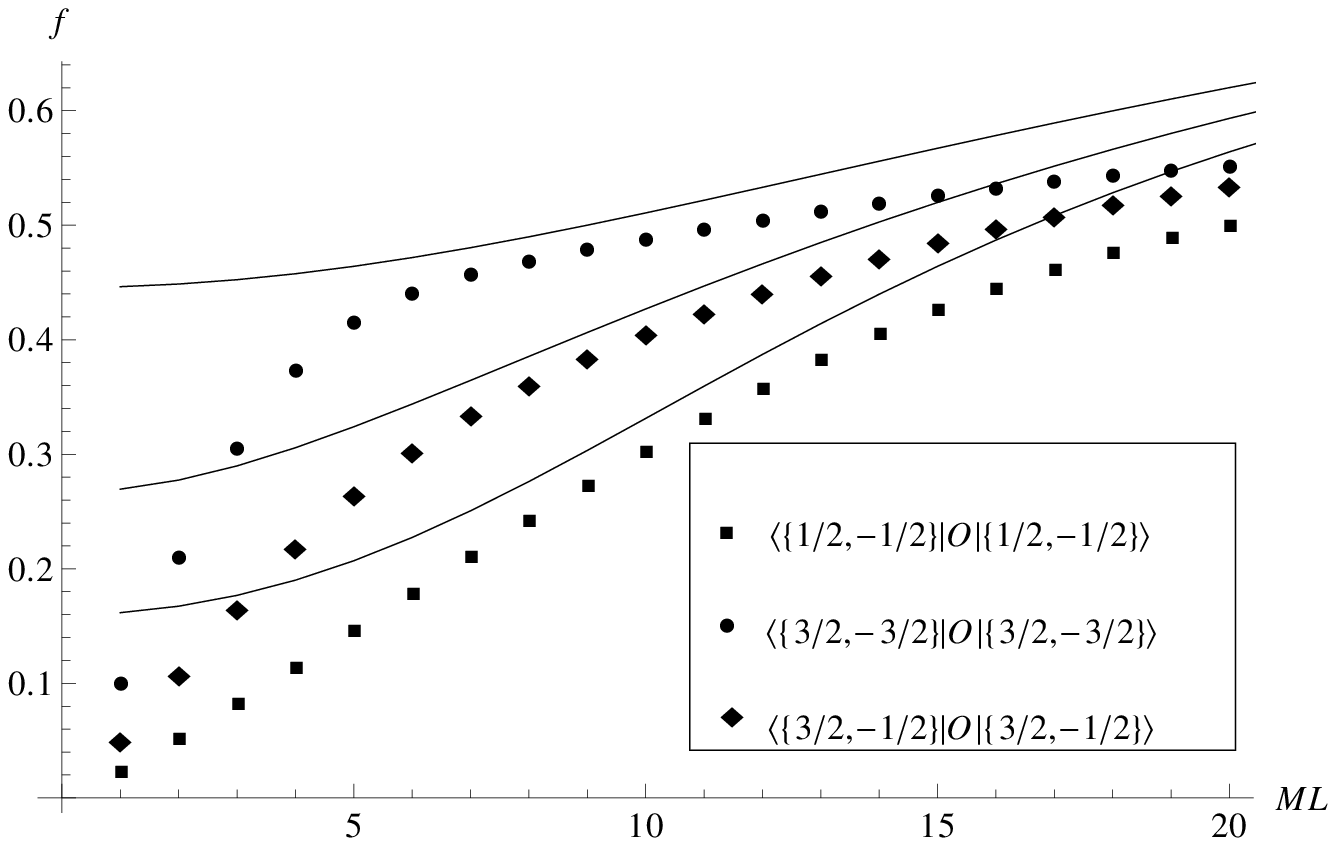}\tabularnewline
$\xi=2/7$\tabularnewline
\includegraphics[width=0.65\textwidth]{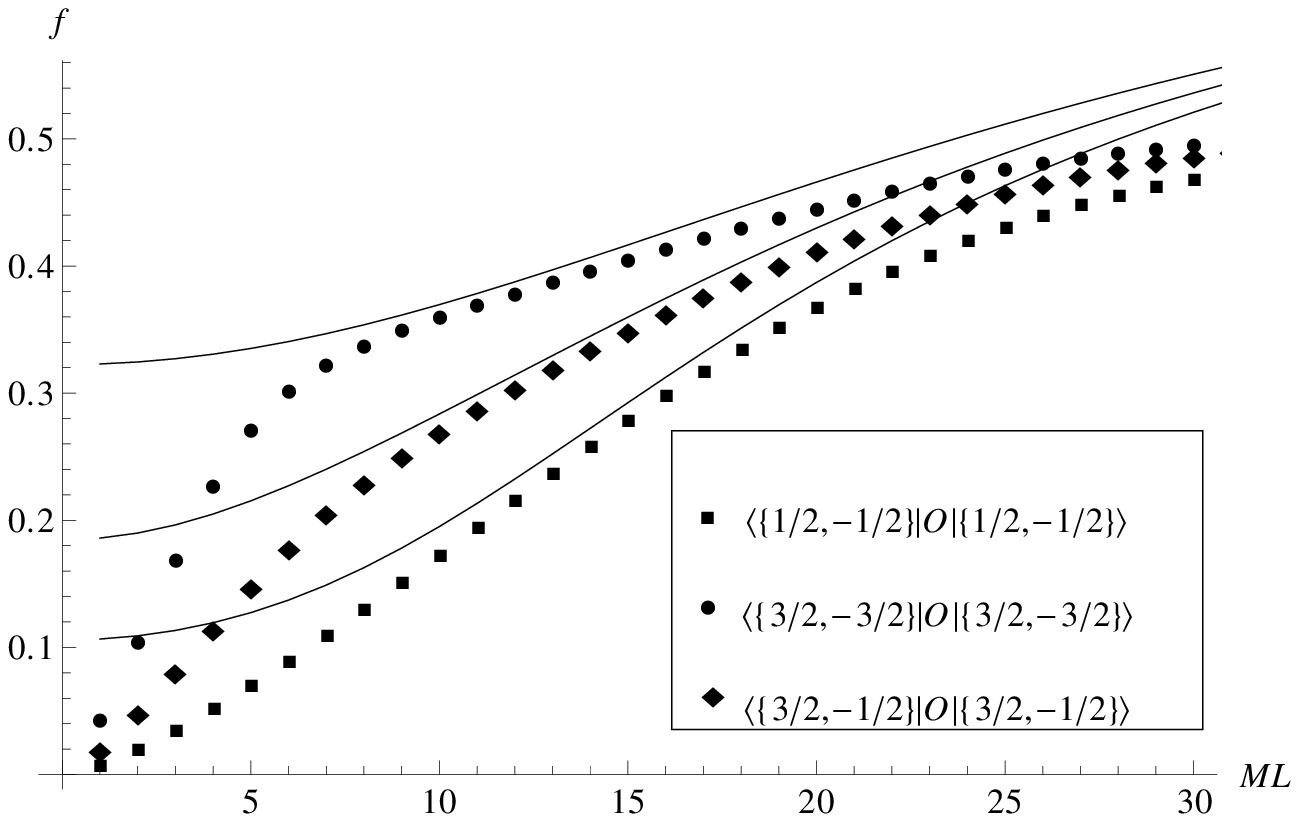}\tabularnewline
$\xi=50/239$\tabularnewline
\includegraphics[width=0.65\textwidth]{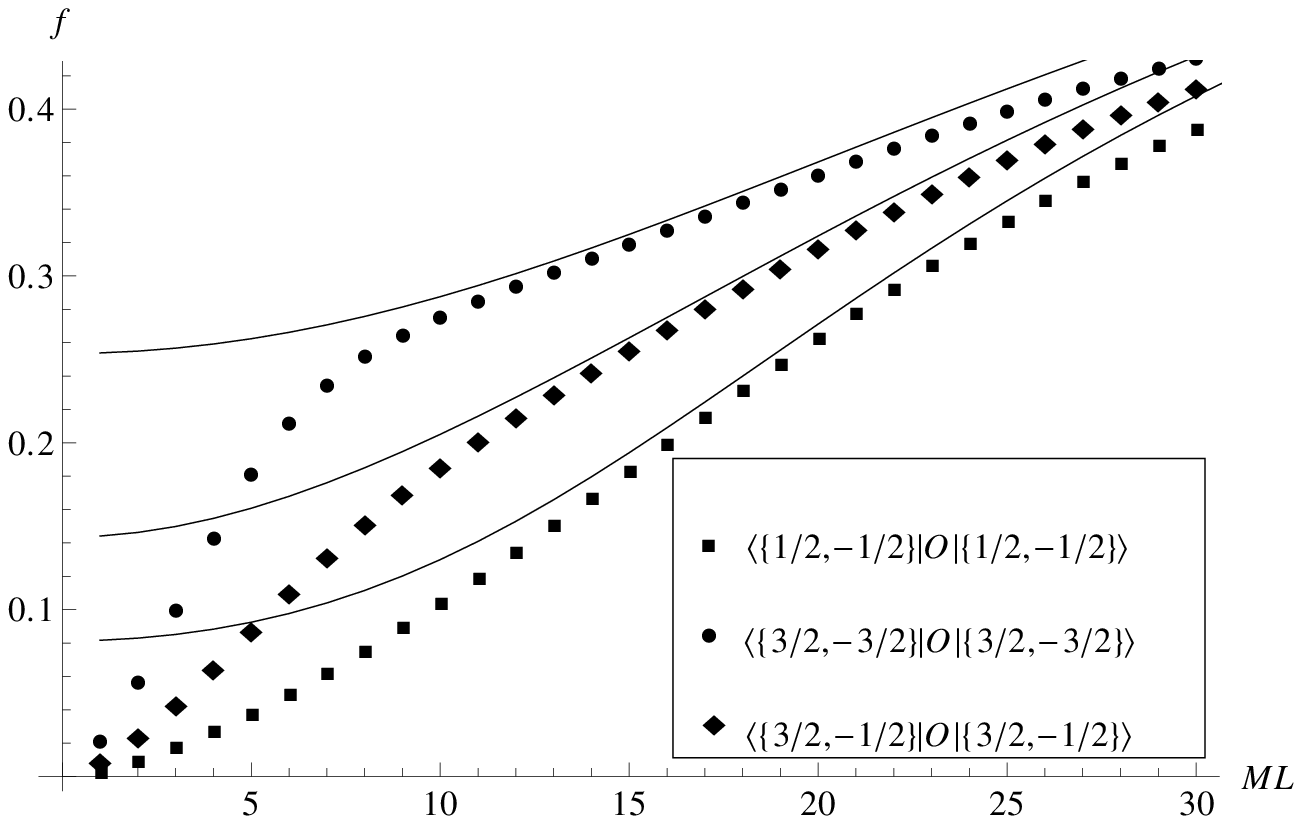}\tabularnewline
$\xi=50/311$ \tabularnewline
\end{tabular}
\par\end{centering}

\caption{\label{fig:q2diag}Diagonal form factors in the $Q=2$ sector}
\end{figure}

\begin{figure}
\begin{centering}
\includegraphics[width=0.6\paperwidth]{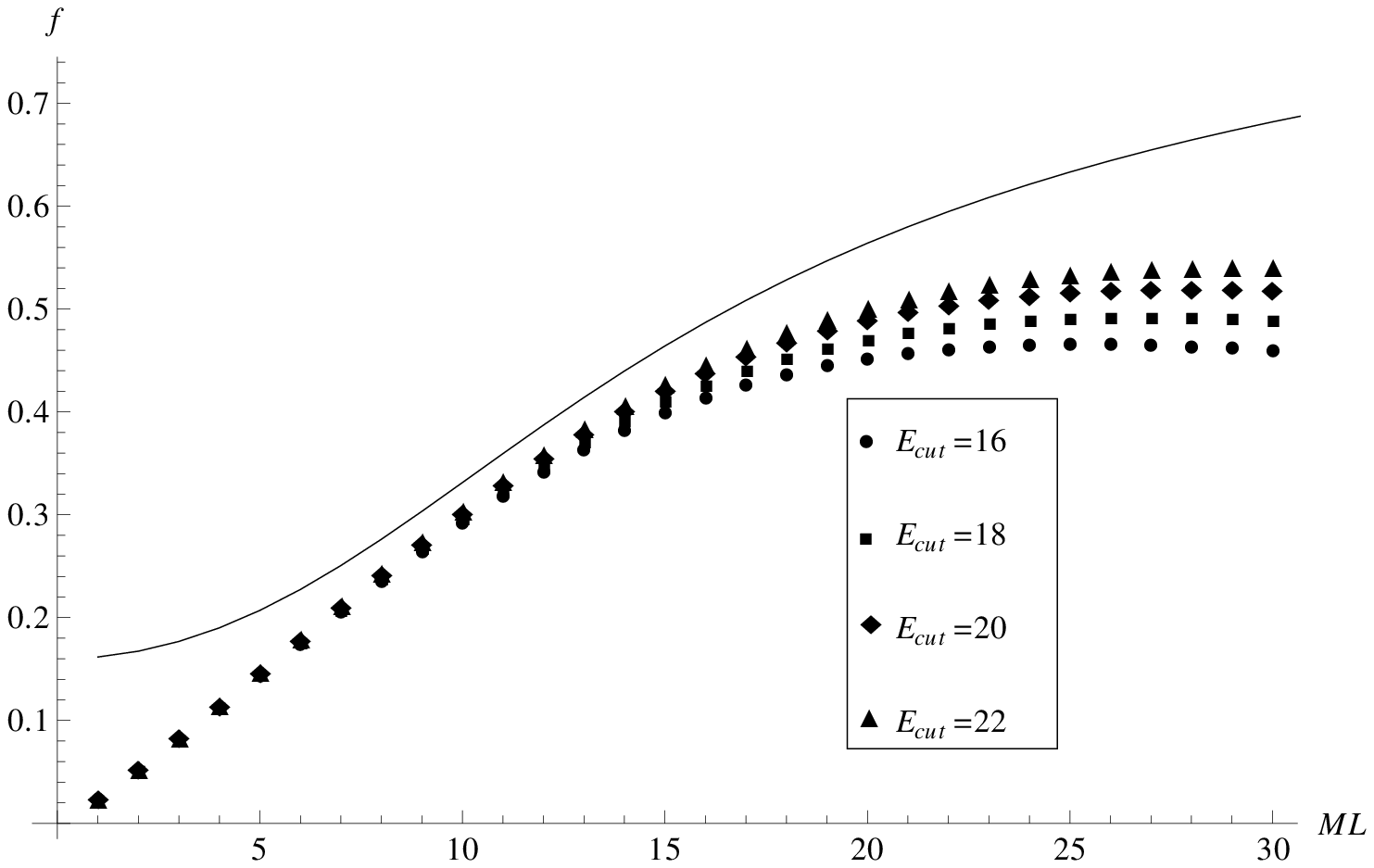}
\par\end{centering}

\caption{\label{fig:Truncation-dependence}Truncation dependence of the diagonal
matrix element $\langle\{1/2,-1/2\}|\mathcal{O}|\{1/2,-1/2\}\rangle$
in the $Q=2$ sector for $\xi=2/7$}
\end{figure}

In addition we can use the soliton-soliton two-particle states. For
off-diagonal matrix elements it is straightforward to use (\ref{eq:genffrelation})
and we obtained a good agreement as demonstrated in fig. \ref{fig:q2nondiag}.
For the diagonal case however, one observes a discrepancy between
the predictions from eqn. (\ref{eq:diaggenrule}) and the numerical
results in fig. \ref{fig:q2diag} which becomes smaller for smaller
values of $\xi$. As illustrated in fig. \ref{fig:Truncation-dependence},
this can be explained by truncation errors, which are indeed improved
by decreasing $\xi$. One can try to extrapolate the truncation dependence;
however, it turns out that it does not fit the theoretically expected
asymptotics derived in \cite{Konik:2007cb}, which means that the
leading order renormalization group behaviour is not yet valid at
the cutoffs considered. Extrapolations reproducing the theoretical
predictions can be found, but for a cut-off dependence which has an
exponent that differs from the predictions of the renormalization
group; in addition, the available range of cutoff values is not sufficient
for a reliable determination of the exponent from the numerical data.
In the conclusions we discuss how the situation can be improved, but
this is out of the scope of the present work.

\subsection{Six-soliton form factors}

\begin{figure}
\begin{centering}
\includegraphics[width=0.6\paperwidth]{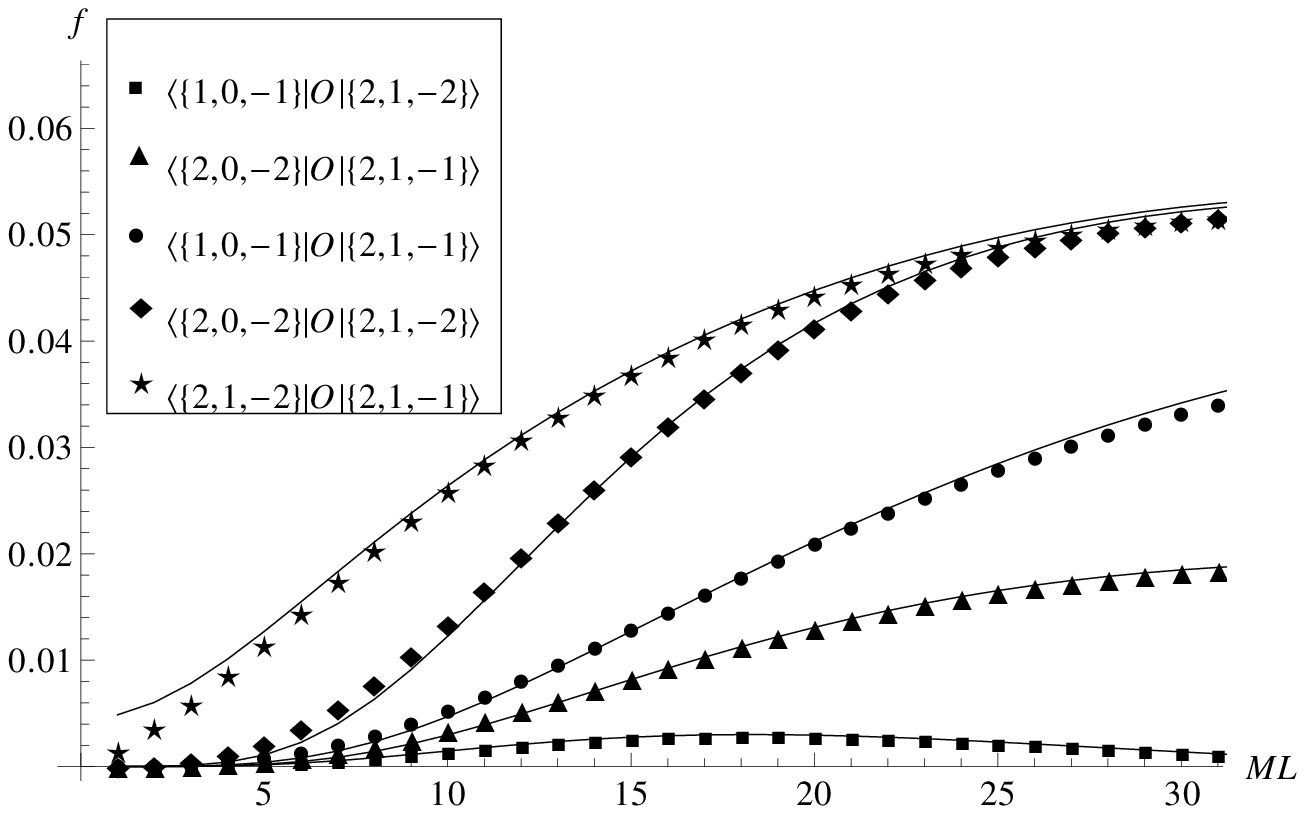}
\par\end{centering}

\caption{\label{fig:q3conn} Off-diagonal form factors in the $Q=3$ sector
for $\xi=50/239$}
\end{figure}

We tested six-soliton form factors by comparing the predictions from
eqn. (\ref{eq:genffrelation}) to off-diagonal matrix elements between
states composed of three solitons (and no antisolitons). The agreement
is again very convincing, as demonstrated in fig. \ref{fig:q3conn}.
For diagonal matrix elements, we noticed similar discrepancies as
in the case of diagonal four-soliton form factors; however, the truncation
dependence proved to be much worse in this case, so while the results
were qualitatively consistent, they were not as good as for the four-soliton
case.

\begin{figure}
\begin{centering}
\includegraphics[width=0.6\paperwidth]{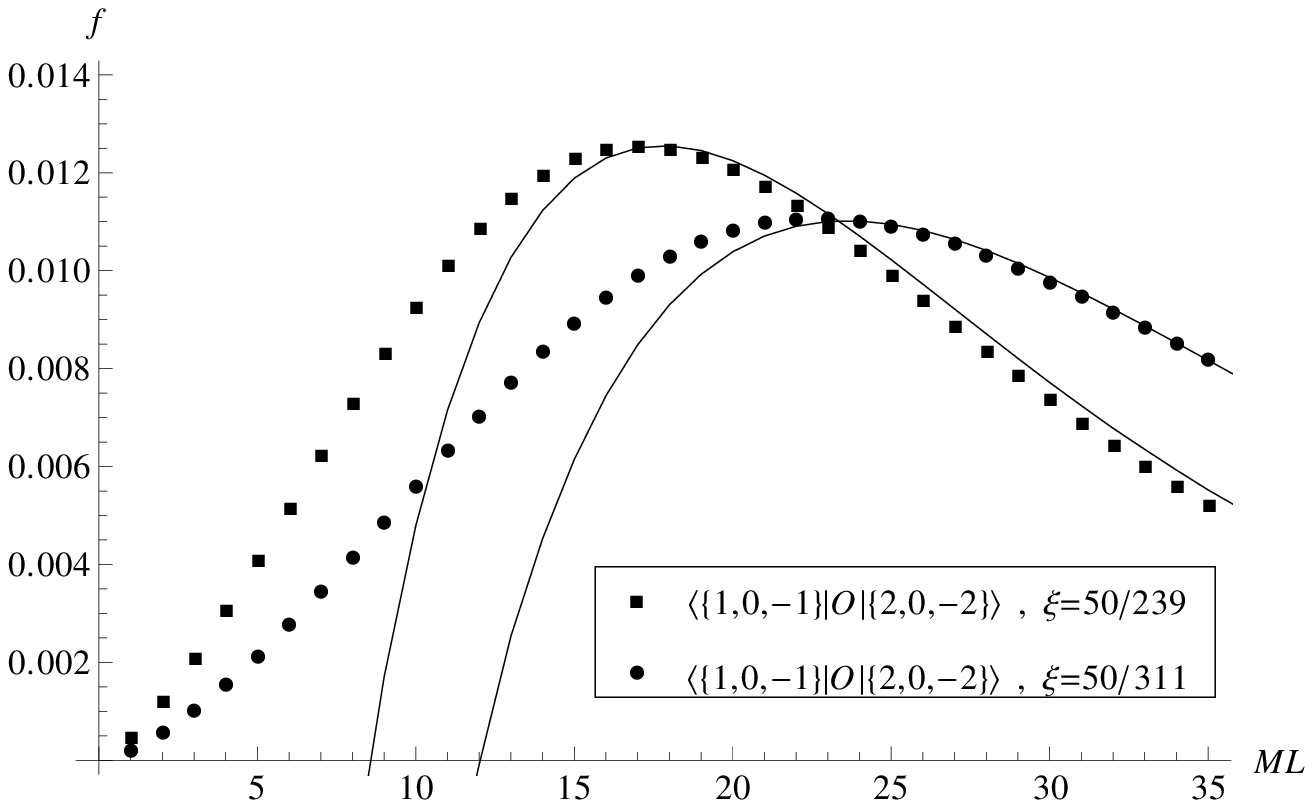}
\par\end{centering}

\caption{\label{fig:zeromom} Form factors with zero-momentum disconnected
pieces in the $Q=3$ sector}
\end{figure}

In addition, for this case there is an interesting new possibility
of having disconnected parts originating from particles with exactly
zero rapidity, described by eqn. (\ref{eq:zeromom}). For these matrix
elements we get a very convincing agreement again, as demonstrated
in fig. \ref{fig:zeromom}.


\section{Conclusions and outlook\label{sec:Conclusions-and-outlook}}

In this work we compared the conjectured exact soliton form factors
of sine-Gordon theory, obtained from the bootstrap, to finite volume
matrix elements given by the truncated conformal space approach. 

For non-diagonal matrix elements we find excellent agreement between
the numerical results and theoretical expectations, both for four-soliton
and six-soliton form factors. For diagonal matrix elements we found
some discrepancy similar to the one noticed in \cite{Feher:2011aa}
for four-breather form factors. This discrepancy tends to decrease
for smaller value of $\xi$ (or equivalently $\beta$) and we argued
that it can be attributed to truncation effects. Similar effects were
observed for boundary form factors in \cite{Lencses:2011ab} and based
on the accumulated data we are inclined to think that there is something
special about cutoff dependence of diagonal matrix elements.

Convergence of the TCSA can be improved by renormalization group methods
\cite{Feverati:2006ni,Konik:2007cb,Giokas:2011ix}; as we discussed
in subsection \ref{sub:Results-for-four-particle} it turns out that
the TCSA data are not yet in the regime where the leading RG behaviour
is applicable, and the extrapolation fits are not reliable enough
to determine the exponent of the cutoff dependence. This can be helped
by applying the numerical RG technique proposed in \cite{Konik:2007cb};
however, developing a systematic program for that takes a substantial
amount of effort and time, and work in this direction has just started.
One can also extend the domain of comparison by improving the theoretical
description for smaller volumes (where truncation errors are negligible)
by describing exponential finite size effects. 

Aside from the above-mentioned technical issues, there is a crucial
missing piece, namely the description of disconnected pieces for states
in which the scattering is non-diagonal, i.e. an extension of formulae
(\ref{eq:diaggenrule}, \ref{eq:zeromom}) to the general case. The
work aimed at resolving this issue is in progress, and the developments
in this paper are also useful in preparing a testing ground for future
theoretical conjectures. Once this final piece is in place, it will
be possible to use the systematic formalism developed in \cite{Pozsgay:2010cr}
for the form factor expansion of finite temperature correlators to
evaluate correlators in field theories with non-diagonal scattering,
such as sine-Gordon theory or the O(3) nonlinear $\sigma$-model. 

\appendix

\section{\label{sec:Explicit-formulae-for-FF} Explicit formulae for the soliton
form factors}

Let us denote the form factor functions defined by Lukyanov by
\[
\mathcal{F}_{\sigma_{2n}\dots\sigma_{1}}^{(a)}(\theta_{2n},\dots,\theta_{1})=\mathcal{G}_{a}(\beta)\langle\langle Z_{\sigma_{2n}}(\theta_{2n})\dots Z_{\sigma_{1}}(\theta_{1})\rangle\rangle
\]
where $\sigma_{i}=\pm$ and 
\[
\sum_{k=1}^{2n}\sigma_{k}=0
\]
is necessary for the matrix element to be different from zero. The
operators $Z$ are given by
\begin{eqnarray*}
Z_{+}(\theta) & = & \sqrt{i\frac{\mathcal{C}_{2}}{4\mathcal{C}_{1}}}\mathrm{e}^{a\theta}\mathrm{e}^{i\phi(\theta)}\\
Z_{-}(\theta) & = & \sqrt{i\frac{\mathcal{C}_{2}}{4\mathcal{C}_{1}}}\mathrm{e}^{-a\theta}\Bigg\{\mathrm{e}^{\frac{i4\pi^{2}}{\beta^{2}}}\int_{C_{+}}\frac{d\gamma}{2\pi}\mathrm{e}^{(1-2a-8\pi/\beta^{2})(\gamma-\theta)}\mathrm{e}^{-i\bar{\phi}(\gamma)}\mathrm{e}^{i\phi(\theta)}\\
 &  & -\mathrm{e}^{-\frac{i4\pi^{2}}{\beta^{2}}}\int_{C_{-}}\frac{d\gamma}{2\pi}\mathrm{e}^{(1-2a-8\pi/\beta^{2})(\gamma-\theta)}\mathrm{e}^{i\phi(\theta)}\mathrm{e}^{-i\bar{\phi}(\gamma)}\Bigg\}
\end{eqnarray*}
The averages are computed by the multiplicative Wick theorem (valid
for exponential operators) using
\begin{eqnarray*}
\langle\langle\mathrm{e}^{i\phi(\theta_{2})}\mathrm{e}^{i\phi(\theta_{1})}\rangle\rangle & = & G(\theta_{1}-\theta_{2})\\
\langle\langle\mathrm{e}^{i\phi(\theta_{2})}\mathrm{e}^{i\bar{\phi}(\theta_{1})}\rangle\rangle & = & W(\theta_{1}-\theta_{2})\\
\langle\langle\mathrm{e}^{i\bar{\phi}(\theta_{2})}\mathrm{e}^{i\bar{\phi}(\theta_{1})}\rangle\rangle & = & \bar{G}(\theta_{1}-\theta_{2})
\end{eqnarray*}
The function $G$ is given by the integral representation
\begin{eqnarray*}
G(\theta) & = & i\mathcal{C}_{1}\sinh\left(\frac{\theta}{2}\right)\exp\left\{ \int_{0}^{\infty}\frac{dt}{t}\sinh^{2}\left(\left(1-\frac{i\theta}{\pi}\right)t\right)\frac{\sinh t\left(\xi-1\right)}{\sinh2t\cosh t\sinh t\xi}\right\} \\
 & = & i\mathcal{C}_{1}\sinh\left(\frac{\theta}{2}\right)\prod_{k=1}^{N}g(\theta,\xi,k)^{k}\,\exp\Bigg\{\int_{0}^{\infty}\frac{dt}{t}\mathrm{e}^{-4Nt}\left(1+N-N\,\mathrm{e^{-4t}}\right)\\
 &  & \times\sinh^{2}\left(\left(1-\frac{i\theta}{\pi}\right)t\right)\frac{\sinh t\left(\xi-1\right)}{\sinh2t\cosh t\sinh t\xi}\Bigg\}
\end{eqnarray*}
where the second formula provides an extension for the domain of convergence
of the integral by factorizing out the following pole factors
\begin{eqnarray*}
g(\theta,\xi,k) & = & \frac{\Gamma\left(\frac{(2k+1+\xi)\pi-i\theta}{\pi\xi}\right)\Gamma\left(\frac{2k+1}{\xi}\right)^{2}\Gamma\left(\frac{(2k+1)\pi-i\theta}{\pi\xi}\right)}{\Gamma\left(\frac{2k+\xi}{\xi}\right)^{2}\Gamma\left(\frac{(2k+\xi)\pi-i\theta}{\pi\xi}\right)\Gamma\left(\frac{(2k-2+\xi)\pi+i\theta}{\pi\xi}\right)}\\
 &  & \times\frac{\Gamma\left(\frac{(2k-1)\pi+i\theta}{\pi\xi}\right)\Gamma\left(\frac{2k-1+\xi}{\xi}\right)^{2}\Gamma\left(\frac{(2k-1+\xi)\pi+i\theta}{\pi\xi}\right)}{\Gamma\left(\frac{2k}{\xi}\right)^{2}\Gamma\left(\frac{(2k+2)\pi-i\theta}{\pi\xi}\right)\Gamma\left(\frac{2k\pi+i\theta}{\pi\xi}\right)}
\end{eqnarray*}
and is independent of $N$,
\begin{eqnarray*}
\mathcal{C}_{1} & = & G(-i\pi)=\exp\left\{ -\int_{0}^{\infty}\frac{dt}{t}\frac{\sinh^{2}(t/2)\sinh(t(\xi-1))}{\sinh(2t)\cosh(t)\sinh(t\xi)}\right\} \\
\mathcal{C}_{2} & = & \exp\left\{ 4\int_{0}^{\infty}\frac{dt}{t}\frac{\sinh^{2}(t/2)\sinh(t(\xi-1))}{\sinh(2t)\sinh(t\xi)}\right\} 
\end{eqnarray*}
and
\begin{eqnarray*}
W(\theta) & = & \frac{1}{G(\theta+i\pi/2)G(\theta-i\pi/2)}\\
 & = & -\frac{2}{\cosh\theta}\prod_{k=1}^{N}\frac{\Gamma\left(\frac{(2k-5/2+\pi)\xi+i\theta}{\pi\xi}\right)\Gamma\left(\frac{(2k-1/2)\pi-i\theta}{\pi\xi}\right)\Gamma\left(\frac{2k-1/2}{\xi}\right)^{2}}{\Gamma\left(1+\frac{2k-3/2}{\xi}\right)^{2}\Gamma\left(\frac{(2k+1/2)\pi-i\theta}{\pi\xi}\right)\Gamma\left(\frac{(2k-3/2)\pi+i\theta}{\pi\xi}\right)}\\
 &  & \times\exp\left\{ -2\int_{0}^{\infty}\frac{dt}{t}\mathrm{e}^{-4Nt}\sinh^{2}\left(\left(1-\frac{i\theta}{\pi}\right)t\right)\frac{\sinh t\left(\xi-1\right)}{\sinh2t\sinh t\xi}\right\} \\
\bar{G}(\theta) & = & \frac{1}{W(\theta+i\pi/2)W(\theta-i\pi/2)}\\
 & = & -\frac{\mathcal{C}_{2}}{4}\xi\sinh\frac{\theta+i\pi}{\xi}\sinh\theta
\end{eqnarray*}
where, again, the integral formula is eventually independent of the
natural number $N$; it provides a representation which converges
faster numerically and is valid further away from the real $\theta$
axis with increasing $N$. The contours in the integrals are such
that the “principal poles” of the $W$-functions are always between
the contour and the real line, where the “principal pole” of $W(x)$
is the one located at $x=-i\pi/2$.

The integral representation can be evaluated in a closed form at the
free fermion point $\xi=1$ and also for the two-particle case when
$a$ is either integer or half-integer \cite{Lukyanov:1997bp}. Here
we only quote the case needed in the text:
\begin{equation}
\mathcal{F}_{\pm\mp}^{1}(\theta)=\mathcal{G}_{1}(\beta)\,\frac{G(\theta)}{G(-i\pi)}\cot\left(\frac{\pi\xi}{2}\right)\frac{4i\cosh\left(\frac{\theta}{2}\right)\mathrm{e}^{\mp\frac{\theta+i\pi}{2\xi}}}{\xi\sinh\left(\frac{\theta+i\pi}{\xi}\right)}\label{eq:twossff}
\end{equation}
The numerical evaluation of the integral representation is rather
involved; the details are given in \cite{Palmai:2011nb}.

\subsubsection*{Acknowledgments}

This work was partially supported by the Hungarian OTKA grant K75172.\\

\subsubsection*{Dedication}

This paper is dedicated to Zalán Horváth, my former MSc and PhD supervisor
and long-time mentor, who recently passed away. It was under his supervision
that I learned about the form factor bootstrap \cite{Horvath:1994rg,Horvath:1996tv},
more specifically about the sine-Gordon form factors studied in this
work. He was interested in finding a direct way to establish that
the exact form factors really give a solution to the quantum field
theory dynamics, which is one of the outcomes of this paper and much
of my recent work.\\
~

~~~~~~~~~~~~~~~~~~~~~~~~~~~~~~~~~~~~~~~~~~~~~~~~~~~~~~~~~~~~~~~~~~~~~~~~~~~~~~~~~~~~~Gábor
Takács

\bibliographystyle{utphys}
\bibliography{sgsolfftcsa}

\providecommand{\href}[2]{#2}\begingroup\raggedright\begin{thebibliography}{10}

\bibitem{zam-zam}
A.~B. Zamolodchikov and A.~B. Zamolodchikov, ``{Factorized S-matrices in two
  dimensions as the exact solutions of certain relativistic quantum field
  models},''
\href{http://dx.doi.org/10.1016/0003-4916(79)90391-9}{{\em Annals Phys.}
  {\bfseries 120} (1979) 253--291}.

\bibitem{Mussardo:1992uc}
G.~Mussardo, ``{Off critical statistical models: Factorized scattering theories
  and bootstrap program},''
\href{http://dx.doi.org/10.1016/0370-1573(92)90047-4}{{\em Phys. Rept.}
  {\bfseries 218} (1992) 215--379}.

\bibitem{Karowski:1978vz}
M.~Karowski and P.~Weisz, ``{Exact Form-Factors in (1+1)-Dimensional Field
  Theoretic Models with Soliton Behavior},''
\href{http://dx.doi.org/10.1016/0550-3213(78)90362-0}{{\em Nucl. Phys.}
  {\bfseries B139} (1978) 455}.

\bibitem{Kirillov:1987jp}
A.~N. Kirillov and F.~A. Smirnov, ``{A representation of the current algebra
  connected with the SU(2) invariant Thirring model},''
\href{http://dx.doi.org/10.1016/0370-2693(87)90908-7}{{\em Phys. Lett.}
  {\bfseries B198} (1987) 506--510}.

\bibitem{Smirnov:1992vz}
F.~A. Smirnov, ``{Form-factors in completely integrable models of quantum field
  theory},''
{\em Adv. Ser. Math. Phys.} {\bfseries 14} (1992) 1--208.

\bibitem{Bajnok:2006ze}
Z.~Bajnok, L.~Palla, and G.~Takacs, ``{On the boundary form factor program},''
  \href{http://dx.doi.org/10.1016/j.nuclphysb.2006.05.019}{{\em Nucl. Phys.}
  {\bfseries B750} (2006) 179--212},
\href{http://arxiv.org/abs/hep-th/0603171}{{\ttfamily arXiv:hep-th/0603171}}.

\bibitem{Takacs:2008je}
G.~Takacs, ``{Form factors of boundary exponential operators in the sinh-Gordon
  model},'' \href{http://dx.doi.org/10.1016/j.nuclphysb.2008.01.025}{{\em Nucl.
  Phys.} {\bfseries B801} (2008) 187--206},
\href{http://arxiv.org/abs/0801.0962}{{\ttfamily arXiv:0801.0962}}.

\bibitem{Cardy:1990pc}
J.~L. Cardy and G.~Mussardo, ``{Form-factors of descendent operators in
  perturbed conformal field theories},''
\href{http://dx.doi.org/10.1016/0550-3213(90)90452-J}{{\em Nucl.Phys.}
  {\bfseries B340} (1990) 387--402}.

\bibitem{Koubek:1993ke}
A.~Koubek and G.~Mussardo, ``{On the operator content of the sinh-Gordon
  model},'' \href{http://dx.doi.org/10.1016/0370-2693(93)90554-U}{{\em Phys.
  Lett.} {\bfseries B311} (1993) 193--201},
\href{http://arxiv.org/abs/hep-th/9306044}{{\ttfamily arXiv:hep-th/9306044}}.

\bibitem{Koubek:1994di}
A.~Koubek, ``{A Method to determine the operator content of perturbed conformal
  field theories},'' \href{http://dx.doi.org/10.1016/0370-2693(94)01680-B}{{\em
  Phys. Lett.} {\bfseries B346} (1995) 275--283},
\href{http://arxiv.org/abs/hep-th/9501028}{{\ttfamily arXiv:hep-th/9501028}}.

\bibitem{Koubek:1994gk}
A.~Koubek, ``{Form-factor bootstrap and the operator content of perturbed
  minimal models},'' \href{http://dx.doi.org/10.1016/0550-3213(94)90368-9}{{\em
  Nucl. Phys.} {\bfseries B428} (1994) 655--680},
\href{http://arxiv.org/abs/hep-th/9405014}{{\ttfamily arXiv:hep-th/9405014}}.

\bibitem{Koubek:1994zp}
A.~Koubek, ``{The Space of local operators in perturbed conformal field
  theories},'' \href{http://dx.doi.org/10.1016/0550-3213(94)00560-2}{{\em Nucl.
  Phys.} {\bfseries B435} (1995) 703--734},
\href{http://arxiv.org/abs/hep-th/9501029}{{\ttfamily arXiv:hep-th/9501029}}.

\bibitem{Smirnov:1995jp}
F.~A. Smirnov, ``{Counting the local fields in SG theory},''
  \href{http://dx.doi.org/10.1016/0550-3213(95)00423-P}{{\em Nucl. Phys.}
  {\bfseries B453} (1995) 807--824},
\href{http://arxiv.org/abs/hep-th/9501059}{{\ttfamily arXiv:hep-th/9501059}}.

\bibitem{Delfino:1995zk}
G.~Delfino and G.~Mussardo, ``{The spin-spin correlation function in the
  two-dimensional Ising model in a magnetic field at $T=T_c$},''
  \href{http://dx.doi.org/10.1016/0550-3213(95)00464-4}{{\em Nucl.Phys.}
  {\bfseries B455} (1995) 724--758},
\href{http://arxiv.org/abs/hep-th/9507010}{{\ttfamily arXiv:hep-th/9507010
  [hep-th]}}.

\bibitem{Delfino:2007bt}
G.~Delfino and G.~Niccoli, ``{Isomorphism of critical and off-critical operator
  spaces in two-dimensional quantum field theory},''
  \href{http://dx.doi.org/10.1016/j.nuclphysb.2008.01.019}{{\em Nucl.Phys.}
  {\bfseries B799} (2008) 364--378},
\href{http://arxiv.org/abs/0712.2165}{{\ttfamily arXiv:0712.2165 [hep-th]}}.

\bibitem{Yurov:1990kv}
V.~P. Yurov and A.~B. Zamolodchikov, ``{Correlation functions of integrable 2-D
  models of relativistic field theory. Ising model},''
\href{http://dx.doi.org/10.1142/S0217751X91001660}{{\em Int. J. Mod. Phys.}
  {\bfseries A6} (1991) 3419--3440}.

\bibitem{Zamolodchikov:1990bk}
A.~B. Zamolodchikov, ``{Two point correlation function in scaling Lee-Yang
  model},''
\href{http://dx.doi.org/10.1016/0550-3213(91)90207-E}{{\em Nucl. Phys.}
  {\bfseries B348} (1991) 619--641}.

\bibitem{Zamolodchikov:1986gt}
A.~B. Zamolodchikov, ``{Irreversibility of the Flux of the Renormalization
  Group in a 2D Field Theory},''
{\em JETP Lett.} {\bfseries 43} (1986) 730--732.

\bibitem{Delfino:1996nf}
G.~Delfino, P.~Simonetti, and J.~L. Cardy, ``{Asymptotic factorisation of form
  factors in two- dimensional quantum field theory},''
  \href{http://dx.doi.org/10.1016/0370-2693(96)01035-0}{{\em Phys. Lett.}
  {\bfseries B387} (1996) 327--333},
\href{http://arxiv.org/abs/hep-th/9607046}{{\ttfamily arXiv:hep-th/9607046}}.

\bibitem{Pozsgay:2007kn}
B.~Pozsgay and G.~Takacs, ``{Form factors in finite volume I: form factor
  bootstrap and truncated conformal space},''
  \href{http://dx.doi.org/10.1016/j.nuclphysb.2007.06.027}{{\em Nucl. Phys.}
  {\bfseries B788} (2008) 167--208},
\href{http://arxiv.org/abs/0706.1445}{{\ttfamily arXiv:0706.1445}}.

\bibitem{Pozsgay:2007gx}
B.~Pozsgay and G.~Takacs, ``{Form factors in finite volume II:disconnected
  terms and finite temperature correlators},''
  \href{http://dx.doi.org/10.1016/j.nuclphysb.2007.07.008}{{\em Nucl. Phys.}
  {\bfseries B788} (2008) 209--251},
\href{http://arxiv.org/abs/0706.3605}{{\ttfamily arXiv:0706.3605}}.

\bibitem{Smirnov:1998kv}
F.~A. Smirnov, ``{Quasi-classical study of form factors in finite volume},''
\href{http://arxiv.org/abs/hep-th/9802132}{{\ttfamily arXiv:hep-th/9802132}}.

\bibitem{korepin-slavnov}
V.~E. {Korepin} and N.~A. {Slavnov}, ``{Form Factors in the Finite Volume},''
  \href{http://dx.doi.org/10.1142/S0217979299002769}{{\em International Journal
  of Modern Physics B} {\bfseries 13} (1999) 2933--2941},
  \href{http://arxiv.org/abs/arXiv:math-ph/9812026}{{\ttfamily
  arXiv:math-ph/9812026}}.

\bibitem{Mussardo:2003ji}
G.~Mussardo, V.~Riva, and G.~Sotkov, ``{Finite-volume form factors in
  semiclassical approximation},''
  \href{http://dx.doi.org/10.1016/j.nuclphysb.2003.08.017}{{\em Nucl. Phys.}
  {\bfseries B670} (2003) 464--478},
\href{http://arxiv.org/abs/hep-th/0307125}{{\ttfamily arXiv:hep-th/0307125}}.

\bibitem{Doyon:2006pv}
B.~Doyon, ``{Finite-temperature form factors: A review},'' {\em SIGMA}
  {\bfseries 3} (2007) 011,
\href{http://arxiv.org/abs/hep-th/0611066}{{\ttfamily arXiv:hep-th/0611066}}.

\bibitem{Pozsgay:2009pv}
B.~Pozsgay, ``{Finite volume form factors and correlation functions at finite
  temperature},''
\href{http://arxiv.org/abs/0907.4306}{{\ttfamily arXiv:0907.4306 [hep-th]}}.

\bibitem{Kormos:2007qx}
M.~Kormos and G.~Takacs, ``{Boundary form factors in finite volume},''
  \href{http://dx.doi.org/10.1016/j.nuclphysb.2008.05.003}{{\em Nucl. Phys.}
  {\bfseries B803} (2008) 277--298},
\href{http://arxiv.org/abs/0712.1886}{{\ttfamily arXiv:0712.1886}}.

\bibitem{Feher:2011aa}
G.~Feher and G.~Takacs, ``{Sine-Gordon form factors in finite volume},''
  \href{http://dx.doi.org/10.1016/j.nuclphysb.2011.06.020}{{\em Nucl.Phys.}
  {\bfseries B852} (2011) 441--467},
  \href{http://arxiv.org/abs/1106.1901}{{\ttfamily arXiv:1106.1901 [hep-th]}}.

\bibitem{Takacs:2011nb}
G.~Takacs, ``{Determining matrix elements and resonance widths from finite
  volume: the dangerous $\mu$-terms},''
  \href{http://dx.doi.org/10.1007/JHEP11(2011)113}{{\em JHEP} {\bfseries 1111}
  (2011) 113},
\href{http://arxiv.org/abs/1110.2181}{{\ttfamily arXiv:1110.2181 [hep-th]}}.

\bibitem{Pozsgay:2006wb}
B.~Pozsgay and G.~Takacs, ``{Characterization of resonances using finite size
  effects},'' \href{http://dx.doi.org/10.1016/j.nuclphysb.2006.05.007}{{\em
  Nucl. Phys.} {\bfseries B748} (2006) 485--523},
\href{http://arxiv.org/abs/hep-th/0604022}{{\ttfamily arXiv:hep-th/0604022}}.

\bibitem{Takacs:2009fu}
G.~Takacs, ``{Form factor perturbation theory from finite volume},''
  \href{http://dx.doi.org/10.1016/j.nuclphysb.2009.10.001}{{\em Nucl. Phys.}
  {\bfseries B825} (2010) 466--481},
\href{http://arxiv.org/abs/0907.2109}{{\ttfamily arXiv:0907.2109}}.

\bibitem{Palmai:2011nb}
T.~Palmai, ``{Regularization of multi-soliton form factors in sine-Gordon
  model},'' \href{http://arxiv.org/abs/1111.7086}{{\ttfamily arXiv:1111.7086
  [math-ph]}}.

\bibitem{Essler:2004ht}
F.~H.~L. Essler and R.~M. Konik, ``{Applications of massive integrable quantum
  field theories to problems in condensed matter physics},''
\href{http://arxiv.org/abs/cond-mat/0412421}{{\ttfamily
  arXiv:cond-mat/0412421}}.

\bibitem{Essler:2009zz}
F.~H.~L. Essler and R.~M. Konik, ``{Finite-temperature dynamical correlations
  in massive integrable quantum field theories},''
  \href{http://dx.doi.org/10.1088/1742-5468/2009/09/P09018}{{\em J. Stat.
  Mech.} {\bfseries 0909} (2009) P09018},
\href{http://arxiv.org/abs/0907.0779}{{\ttfamily arXiv:0907.0779
  [cond-mat.str-el]}}.

\bibitem{Pozsgay:2010cr}
B.~Pozsgay and G.~Takacs, ``{Form factor expansion for thermal correlators},''
  \href{http://dx.doi.org/10.1088/1742-5468/2010/11/P11012}{{\em J. Stat.
  Mech.} {\bfseries 1011} (2010) P11012},
\href{http://arxiv.org/abs/1008.3810}{{\ttfamily arXiv:1008.3810}}.

\bibitem{Essler:2007jp}
F.~H.~L. Essler and R.~M. Konik, ``{Finite-temperature lineshapes in gapped
  quantum spin chains},''
  \href{http://dx.doi.org/10.1103/PhysRevB.78.100403}{{\em Phys. Rev.}
  {\bfseries B78} (2008) 100403},
\href{http://arxiv.org/abs/0711.2524}{{\ttfamily arXiv:0711.2524
  [cond-mat.str-el]}}.

\bibitem{Kormos:2010ae}
M.~Kormos and B.~Pozsgay, ``{One-Point Functions in Massive Integrable QFT with
  Boundaries},'' \href{http://dx.doi.org/10.1007/JHEP04(2010)112}{{\em JHEP}
  {\bfseries 1004} (2010) 112},
\href{http://arxiv.org/abs/1002.2783}{{\ttfamily arXiv:1002.2783 [hep-th]}}.

\bibitem{Yurov:1989yu}
V.~P. Yurov and A.~B. Zamolodchikov, ``{Truncated conformal space approach to
  scaling Lee-Yang model},''
\href{http://dx.doi.org/10.1142/S0217751X9000218X}{{\em Int. J. Mod. Phys.}
  {\bfseries A5} (1990) 3221--3246}.

\bibitem{Feverati:1998va}
G.~Feverati, F.~Ravanini, and G.~Takacs, ``{Truncated conformal space at c = 1,
  nonlinear integral equation and quantization rules for multi-soliton
  states},'' \href{http://dx.doi.org/10.1016/S0370-2693(98)00543-7}{{\em Phys.
  Lett.} {\bfseries B430} (1998) 264--273},
\href{http://arxiv.org/abs/hep-th/9803104}{{\ttfamily arXiv:hep-th/9803104}}.

\bibitem{smirnov_ff}
F.~A. Smirnov, ``{Form-factors in completely integrable models of quantum field
  theory},''
{\em Adv. Ser. Math. Phys.} {\bfseries 14} (1992) 1--208.

\bibitem{Lukyanov:1993pn}
S.~L. Lukyanov, ``{Free field representation for massive integrable models},''
  \href{http://dx.doi.org/10.1007/BF02099357}{{\em Commun. Math. Phys.}
  {\bfseries 167} (1995) 183--226},
\href{http://arxiv.org/abs/hep-th/9307196}{{\ttfamily arXiv:hep-th/9307196}}.

\bibitem{Lukyanov:1997bp}
S.~L. Lukyanov, ``{Form factors of exponential fields in the sine-Gordon
  model},'' \href{http://dx.doi.org/10.1142/S0217732397002673}{{\em Mod. Phys.
  Lett.} {\bfseries A12} (1997) 2543--2550},
\href{http://arxiv.org/abs/hep-th/9703190}{{\ttfamily arXiv:hep-th/9703190}}.

\bibitem{Babujian:1998uw}
H.~M. Babujian, A.~Fring, M.~Karowski, and A.~Zapletal, ``{Exact form factors
  in integrable quantum field theories: The sine-Gordon model},''
  \href{http://dx.doi.org/10.1016/S0550-3213(98)00737-8}{{\em Nucl. Phys.}
  {\bfseries B538} (1999) 535--586},
\href{http://arxiv.org/abs/hep-th/9805185}{{\ttfamily arXiv:hep-th/9805185}}.

\bibitem{Babujian:2001xn}
H.~Babujian and M.~Karowski, ``{Exact form factors in integrable quantum field
  theories: The sine-Gordon model. II},''
  \href{http://dx.doi.org/10.1016/S0550-3213(01)00551-X}{{\em Nucl. Phys.}
  {\bfseries B620} (2002) 407--455},
\href{http://arxiv.org/abs/hep-th/0105178}{{\ttfamily arXiv:hep-th/0105178}}.

\bibitem{Lukyanov:1996jj}
S.~L. Lukyanov and A.~B. Zamolodchikov, ``{Exact expectation values of local
  fields in quantum sine-Gordon model},''
  \href{http://dx.doi.org/10.1016/S0550-3213(97)00123-5}{{\em Nucl. Phys.}
  {\bfseries B493} (1997) 571--587},
\href{http://arxiv.org/abs/hep-th/9611238}{{\ttfamily arXiv:hep-th/9611238}}.

\bibitem{Zamolodchikov:1995xk}
A.~B. Zamolodchikov, ``{Mass scale in the sine-Gordon model and its
  reductions},''
\href{http://dx.doi.org/10.1142/S0217751X9500053X}{{\em Int. J. Mod. Phys.}
  {\bfseries A10} (1995) 1125--1150}.

\bibitem{Pozsgay:2008bf}
B.~Pozsgay, ``{Luscher's mu-term and finite volume bootstrap principle for
  scattering states and form factors},''
  \href{http://dx.doi.org/10.1016/j.nuclphysb.2008.04.021}{{\em Nucl. Phys.}
  {\bfseries B802} (2008) 435--457},
\href{http://arxiv.org/abs/0803.4445}{{\ttfamily arXiv:0803.4445 [hep-th]}}.

\bibitem{Feverati:2006ni}
G.~Feverati, K.~Graham, P.~A. Pearce, G.~Z. Toth, and G.~Watts, ``{A
  Renormalisation group for TCSA},''
  \href{http://arxiv.org/abs/hep-th/0612203}{{\ttfamily arXiv:hep-th/0612203
  [hep-th]}}.

\bibitem{Konik:2007cb}
R.~M. Konik and Y.~Adamov, ``{A Numerical Renormalization Group for Continuum
  One-Dimensional Systems},''
  \href{http://dx.doi.org/10.1103/PhysRevLett.98.147205}{{\em Phys. Rev. Lett.}
  {\bfseries 98} (2007) 147205},
  \href{http://arxiv.org/abs/cond-mat/0701605}{{\ttfamily
  arXiv:cond-mat/0701605 [cond-mat.str-el]}}.

\bibitem{Giokas:2011ix}
P.~Giokas and G.~Watts, ``{The renormalisation group for the truncated
  conformal space approach on the cylinder},''
  \href{http://arxiv.org/abs/1106.2448}{{\ttfamily arXiv:1106.2448 [hep-th]}}.

\bibitem{Lencses:2011ab}
M.~Lencses and G.~Takacs, ``{Breather boundary form factors in sine-Gordon
  theory},'' \href{http://dx.doi.org/10.1016/j.nuclphysb.2011.07.010}{{\em
  Nucl.Phys.} {\bfseries B852} (2011) 615--633},
\href{http://arxiv.org/abs/1106.1902}{{\ttfamily arXiv:1106.1902 [hep-th]}}.

\bibitem{Horvath:1994rg}
Z.~Horvath and G.~Takacs, ``{Free field representation for the O(3) nonlinear
  sigma model and bootstrap fusion},''
  \href{http://dx.doi.org/10.1103/PhysRevD.51.2922}{{\em Phys. Rev.} {\bfseries
  D51} (1995) 2922--2932},
\href{http://arxiv.org/abs/hep-th/9501006}{{\ttfamily arXiv:hep-th/9501006}}.

\bibitem{Horvath:1996tv}
Z.~Horvath and G.~Takacs, ``{Form-factors of the sausage model obtained with
  bootstrap fusion from sine-Gordon theory},''
  \href{http://dx.doi.org/10.1103/PhysRevD.53.3272}{{\em Phys. Rev.} {\bfseries
  D53} (1996) 3272--3284},
\href{http://arxiv.org/abs/hep-th/9601040}{{\ttfamily arXiv:hep-th/9601040}}.

\end{thebibliography}\endgroup
 \bibliographystyle{utphys}
\end{document}